\documentclass[12pt]{article}
\pdfoutput=1
\usepackage{amsmath,amsfonts,graphicx,color,bbm,tikz,bm,setspace}
\usepackage{comment}
\usepackage[nosort]{cite}
\usepackage{subfigure}
\usetikzlibrary{calc,positioning}
\usetikzlibrary{patterns,arrows,decorations.pathreplacing}
\usepackage{caption}
\usepackage{ulem}
\tikzset{>=stealth}
\usepackage{lipsum}
\usepackage{tabularx}
\usepackage{fullpage}
\usepackage{hyperref}
\usepackage{longtable}
\usepackage{bbold}

\newtheorem{remark}{Remark}[section]
\usepackage{listings}
\usepackage{tcolorbox}

\newcommand{\mathsym}[1]{{}}
\newcommand{\unicode}[1]{{}}

\makeatletter\@addtoreset{equation}{section}\makeatother

\newcommand{\be}{\begin{equation}}
\newcommand{\ee}{\end{equation}}
\def\beq{\begin{equation}}
\def\eeq{\end{equation}}
\newcommand{\bea}{\begin{eqnarray}}
\newcommand{\eea}{\end{eqnarray}}

\newcommand{\tr}{{\rm tr\,}}

\renewcommand{\title}[1]{\vbox{\center\LARGE{#1}}\vspace{3mm}}
\renewcommand{\author}[1]{\vbox{\center{#1}}\vspace{3mm}}

\newcommand{\email}[1]{\vbox{\center\tt#1}\vspace{3mm}}

\begin{document}

\begin{center}
{\large {\bf The Yang-Baxter integrability of the critical Ising chain}}

\author{ Akash Sinha,$^a$ Tinu Justin,$^a$ Pramod Padmanabhan,$^a$ Vladimir Korepin$^{b}$}

{$^a${\it School of Basic Sciences,\\ Indian Institute of Technology, Bhubaneswar, 752050, India}}
\vskip0.1cm
{$^b${\it C. N. Yang Institute for Theoretical Physics, \\ Stony Brook University, New York 11794, USA}}

\email{akash26121999@gmail.com, tinu232000@gmail.com, pramod23phys@gmail.com, vladimir.korepin@stonybrook.edu}

\vskip 0.5cm 

\end{center}


\abstract{
\noindent 
We show that the one dimensional, critical transverse field Ising model is Yang-Baxter integrable. This is done by constructing commuting transfer matrices built out of a $R$-matrix satisfying the Yang-Baxter equation with additive spectral parameters.  The $R$-matrix is non-local, as it is expressed in terms of Majorana fermions. It is also non-regular.  Nevertheless, we show that the quantum inverse scattering method can still be suitably adapted. We then recursively obtain the conserved quantities [in the infinite volume] by the boost operator method. Remarkably, among the conserved charges we also find the Kramers-Wannier duality and other non-invertible symmetries for the periodic transverse field Ising model.
}

\tableofcontents 

\section{Introduction}
\label{sec:Introduction}
The classical two dimensional Ising model \cite{McCoyWu+1973} is known for both its physical relevance and theoretical importance. The former stems from the fact that these systems serve as a simple model for ferromagnetic materials. On the other hand its role as a playground to visualize several theoretical ideas comes from its exact solvability in the absence of a magnetic field\footnote{For results in the presence of a magnetic field see \cite{Fonseca2001IsingFT,Delfino2003IntegrableFT,Henkel1989TheTI,Delfino_1995}.} \cite{Onsager1944,Kaufman1949}. In particular they help understand the nature of phase transitions, the role played by spatial dimensions in the same, and the behavior of correlation functions in phase transitions \cite{McCoy1976,Hecht1967,Yurov1991CORRELATIONFO,Sato:1977uv,McCoy:1977er}. Their continuum versions also help demonstrate the notions of universality and the scaling hypothesis \cite{Sachdev_2011}. It is thus well-known that the Ising model is a highly successful theoretical model that has stood the test of time.

 Our goal in this work is to show the Yang-Baxter integrability of the quantum Ising model at criticality and subsequently compute the conserved quantities of this system. In light of this, it is crucial to trace the history of the development of the two dimensional Ising model that is relevant to this work. We especially focus on the early understanding of quantum integrability formulated in the study of exactly solvable models and then move on to a more modern interpretation which is in line with the Lax formalism and classical integrability. The contrast is essential to understand the contribution of this work.

We begin with the estimation of the critical temperature of this system. Peierls intuitively estimated the critical temperature for the phase transition from the ordered to the disordered phase \cite{Peierls_1936}. This was later made more precise by Kramers and Wannier \cite{KW1941-1,KW1941-2} through the notion of a {\it duality transformation} \cite{Savit1981} that generally relates the partition functions of the system on a given lattice and its dual. This can also be interpreted as a relation between the `high' and `low' temperature expansions of the partition function. This duality, on the hexagonal lattice and its dual triangular lattice, is satisfied when the coupling parameters of the Ising model satisfy an additional constraint called the {\it star-triangle relation} (See Eq. 6.4.7 of \cite{Baxter:1982zz} for explanations and history, this relation was also mentioned in \cite{Wannier1945,Onsager1944, Houtappel1950} and see \cite{au1989onsager} for a review)\footnote{The star-triangle relations can also be viewed as the early forms of the Yang-Baxter equation. It was used to solve many other models including the chiral Potts model \cite{AUYANG1987219,MCCOY19879,BAXTER1988138}.}\footnote{The star-triangle relation also plays an important role in computing several other quantities of physical interest in the Ising model. For a thorough account of these techniques see standard textbooks on this subject \cite{Baxter:1982zz,Mussardo-book}.}. When the Boltzmann weights describing the dynamics are written in operator form, the star-triangle relation becomes the braided form of the {\it Yang-Baxter equation} (YBE) (See Eq. 6.4.25 of \cite{Baxter:1982zz} for the derivation)\footnote{{Recent works that derive $R$-matrices related to spin models from vertex models can be found in \cite{marcio-1,MARCIO2024116610}.}}. This relation can then be used to show that commutativity of the corresponding transfer matrices (See Eq. 6.4.31 of \cite{Baxter:1982zz} and also \cite{mills_ascher_jaffee_1971}). Following this one can use this transfer matrix to arrive at the Hamiltonian limit \cite{Susskind1978} of the two dimensional [1+1 spacetime dimensions] Ising model by taking the lattice spacing of the time direction to zero. This results in a rotated version of the Hamiltonian:
\begin{eqnarray}\label{eq:Ising-H}
    H_{\rm TFIM} = -J\sum\limits_{k=1}^N~X_kX_{k+1} - h\sum\limits_{k=1}^N~Z_k,
\end{eqnarray}
as shown in Chapter 9 of \cite{Mussardo-book}. This is also known as the {\it transverse field Ising model} (TFIM). Criticality occurs when $J=h$.
It should be noted that the classical two dimensional Ising model is only weakly equivalent to the quantum TFIM, in the sense that the two models share critical exponents. The transfer matrix of the classical model does not commute with the quantum Hamiltonian of TFIM \cite{Plischke1970,Elliott1970}. On the other hand, strong equivalence is discussed by M. Suzuki \cite{Suzuki1971-1} where he proves that the transfer matrix of the two dimensional Ising model without a magnetic field commutes with the quantum Hamiltonian of the $XY$ model with a magnetic field. This implies that the two operators share eigenvectors and Suzuki shows that the eigenvector of the transfer matrix with maximal eigenvalue\footnote{Eigenvalues are trace class functions and high and low temperature expansions of partition functions can be written in terms of them. The leading order term is the maximal eigenvalue and dominates in such expansions.} coincides with the ground state of the $XY$ model with a magnetic field \cite{Suzuki1971-1}, thus establishing the strong equivalence between the two systems. He also generalizes these ideas to other systems and higher dimensions in \cite{Suzuki1976-2}. This fact is another motivation for our work as we search for a transfer matrix that exactly contains the TFIM and also commutes with it. 

The above early formulation of the YBE was reinterpreted {\it via} the {\it quantum inverse scattering method} (QISM) by the Leningrad school [L. D. Faddeev, L. A. Takhtajan, E. K. Sklyanin, N. Y. Reshetikhin, F. Smirnov and P.P. Kulish {\it et. al.}] \cite{faddeev1996algebraic,Korepin1993QuantumIS,Sklyanin1982QuantumVO,Sklyanin1979QuantumIP}. This is a purely operator theoretic formulation that ended up giving more algebraic structure to the YBE that includes the discovery of quantum groups \cite{Drinfeld1988,Majid_1995}. This approach starts with a solution of the YBE, also known as the $R$-matrix. Treating the $R$-matrix as a Lax operator we can canonically construct a monodromy matrix on a closed one dimensional lattice or chain. In most cases this matrix automatically satisfies a YBE-like equation called the $RTT$ relation [historically first name was intertwining relation]. This equation then implies the existence of an infinite number of mutually commuting operators. This is true for all invertible $R$-matrices. When one of the mutually commuting operators is interpreted as the Hamiltonian, the existence of the $R$-matrix implies the integrability of this Hamiltonian or in other words it means that this Hamiltonian has many conserved quantities. Techniques like the {\it boost operator method} \cite{loebbert2016lectures} have been developed to construct these conserved charges for special types of $R$-matrices that are regular and have an additive dependence on the spectral parameters. The QISM does not stop here. The operator form of the YBE then lays the foundation for the {\it algebraic Bethe ansatz} \cite{slavnov2019algebraicbetheansatz}, a systematic technique to obtain the spectrum of the Hamiltonian and also the correlation functions of the system \cite{Korepin1993QuantumIS,Korepin1984CorrelationFO,Korepin1994CorrelationFO,Izergin1984TheQI,Maillet2006CorrelationFO}. Thus this helps study all the physical properties of the system including critical exponents and phase transitions.  

The QISM technique has been well developed for some of the famous spin chains such as the $XXX$, $XXZ$, $XYZ$ \cite{Faddeev1984SpectrumAS,Kirillov1986ExactSO,Kirillov1987ExactSO,Takhtadzhan1979THEQM} and other models such as the chiral Potts model \cite{AuYang1996TheMF,AuYang2016About3Y,Perk2015TheEH}, the Hubbard model \cite{Shiroishi1995YangBaxterEF,Shastry1986ExactIO,Shastry1988DecoratedSR,esslerfrahmgohmannklumperkorepin2005} etc. However there is still a gap in the possibility of using QISM to study the Ising Hamiltonian in \eqref{eq:Ising-H}. This leads us to the following question :  
 
{\it Is there a R-matrix, satisfying the Yang-Baxter equation, that can be used to construct commuting transfer matrices that contains the one dimensional quantum Ising model [TFIM] at criticality as a conserved operator?\footnote{{Fermionic realizations of Boltzmann weights can lead to alternate derivations of the TFIM \cite{sedrakyan-1}. A quantum $L$-operator satisfying the $RLL$ relation was recently found in \cite{PronkoSyrygina2024}.}} If yes, can we also find the higher conserved quantities of the TFIM using this $R$-matrix?}

There are several reasons to expect the existence of such a $R$-matrix, from the point of view of general theoretical principles and also from a more technical side. The former arises from the observation that the Ising Hamiltonian \eqref{eq:Ising-H} can be rewritten in terms of fermionic operators acting on a Fock space. In this language it is indeed a {\it free theory} as the `local' terms now are quadratic in the fermionic creation and annihilation operators \cite{Lieb1964}. This makes the model exactly solvable on this space. Now we expect exactly solvable models to also be quantum integrable and in particular Yang-Baxter integrable\footnote{See \cite{Caux_2011} for discussions on more broader notions of quantum integrability.}\footnote{Yang-Baxter equation is a sufficient condition for integrability but not necessary. So not all exactly solvable models are necessarily Yang-Baxter integrable. There are some interesting examples like the Rabi model that is exactly solvable and integrable in a non-Yang-Baxter way \cite{Braak2011}. However it is Yang-Baxter integrable only at certain points of its parameter space \cite{Batchelor_2015}. } \cite{YangCN1967,BAXTER1972193}. This requires the existence of the proposed $R$-matrix. Though the model is exactly solvable in the entire parameter space of $(J,h)$, integrability is expected only at the critical point $J=h$. There is no general principle saying that critical models are also integrable. Nevertheless there are several instances where this is true, for example the $XXX$ spin chain or the Heisenberg model. Also the fact that correlation functions in conformal theories for critical systems can be purely obtained from the symmetries of the theory alone suggests a relation to Yang-Baxter integrable systems. This is because one would imagine that the conserved quantities obtained from a $R$-matrix would be enough to obtain the correlation functions of an integrable system \cite{BERNARD_1993}.

From the more technical side we know that correlation functions of the exactly solvable/integrable $XY$ chain is related to those of the TFIM \cite{ITS1990752,MCCOY198335,Barouch1971,Its1993}. As the $XY$ model is obtainable through a $R$-matrix we would expect the same for the TFIM, driving another motivation for the construction of the $R$-matrix for TFIM at criticality. 


With all this motivation, we will construct a $R$-matrix that answers the above questions. We find both spectral parameter dependent and spectral parameter independent [constant] Yang-Baxter operators that are required to develop integrable models. This $R$-matrix is unusual in the sense that it is not local like the $R$-matrices for other well known integrable spin chains such as the $XXX$ model and its variants. The reason being that it is constructed using Majorana fermions that are inherently non-local as explained in detail in Sec. \ref{sec:majorana-solution}.  They are also not regular in the traditional sense as it reduces to a permutation-like operator for special values of the spectral parameters. Thus this $R$-matrix challenges the scope of the definition of regularity. We will then treat these $R$-matrices like Lax operators that form a monodromy matrix satisfying a $RTT$ relation. Following this we construct transfer matrices out of the monodromy matrices by tracing out the auxiliary indices. The YBE is then used to prove the commutativity of these transfer matrices in Sec. \ref{sec:commutativity-transfer}. While many of these results are quite straightforward in the case of local $R$-matrices, similar proofs become significantly more involved in the case of non-local $R$-matrices studied here. For this reason all proofs are presented in some detail. These complications also extend to the derivation of the local Hamiltonian [Sec. \ref{sec:critical-Ising}] and other conserved quantities [Sec. \ref{sec:boost}] from this transfer matrix using the suitably modified boost operator method. After the dust settles we arrive at the critical Ising chain with a magnetic field [the TFIM] for both periodic and anti-periodic boundary conditions in Sec. \ref{subsec:Ising-H}. At criticality the Ising chain commutes with the {\it Kramers-Wannier (KW) duality} operator and so we expect this operator to be contained in the expansion of the transfer matrix. This is indeed the case as demonstrated in Sec. \ref{subsec:KW-symmetry}. This is especially interesting given the recent attention that the KW duality has received in the context of topological defects \cite{frohlich2004kramers,aasen2016topological,aasen2020topological} and non-invertible symmetries \cite{shao2023s}. This work lays the foundation for further studies on the quantum Ising model from the perspective of the quantum inverse scattering method. These are briefly touched upon in Sec. \ref{sec:conclusion}. Other future directions are also discussed here.

\subsection*{Summary of results}
Due to use of non-local Majorana $R$-matrices, a number of properties of integrable systems require a derivation as we cannot assume the results for the local counterparts. This makes the paper quite dense. Thus this little section serves as a guide to the main results that a casual reader can easily look up to.
\begin{enumerate}
    \item Solutions for the Yang-Baxter equation form the backbone of this work. The constant Majorana braid operator is given by \eqref{eq:P-Majorana}. This operator resembles a permutation type operator. {The additive $R$-matrix that we construct reduces to this operator when the spectral parameter goes to 0. As this operator is not the usual permutation operator, we consider this $R$-matrix as non-regular, simply because it does not fit the usual definition of regularity.} This spectral parameter dependent braided $R$-matrix can be found in \eqref{eq:R-spec-Majorana-braided}. Two equivalent versions of the non-braided form of the latter is in \eqref{eq:R-matrix-form1-form2}. The non-braided $R$-matrix is used in subsequent sections to show the integrability of the critical Ising chain.
    \item The $RTT$ relation [also known as intertwining relation] underlies the integrability of the Hamiltonians derived from the $R$-matrices. The proof is shown in \eqref{eq:proof-RTT}. Following this the integrability can be proved by tracing out auxiliary space indices. This is detailed in equations \ref{eq:T-odd-even} through \ref{eq:TMcom}.
    \item The transfer matrix constructed out of the Majoranas is central to the rest of the paper. Writing this in different forms sheds light on different aspects of the critical Ising chain. The first form, useful for deriving the Hamiltonian is shown in \eqref{eq:transfer-matrix-lambda0}. This can also be recast as a translation operator on the space of Majoranas as in \eqref{eq:Majorana-translation-unitary}. To relate the latter to the Kramers-Wannier duality it is useful to rewrite this as \eqref{eq:Majorana-translation-KW}.
    \item The local Hamiltonian in terms of the Majorana fermions derived from the transfer matrices is in \eqref{eq:kitaevH-antiperiodic}. This can be related to the Ising model in the spin basis using a slightly altered Jordan-Wigner transform in \eqref{eq:JW-transform}. 
    \item The Kramers-Wannier duality symmetry can be derived from the Majorana transfer matrix as shown in \eqref{eq:KW-duality-operator}.
    \item The boost operator method is generalized to find that the general form of the higher conserved charges is given by \eqref{eq:general-Qp}. This is proved using induction in Sec. \ref{sec:boost}.
\end{enumerate}

\subsection*{A note on notations}
As this work is intended for readers with different expertise we clarify some of the notation used in this work.
\begin{enumerate}
    \item The letter $P$ is reserved for the standard permutation operator. Modified versions of the permutation operator contain additional symbols on $P$.
    \item The solutions of the {non-braided} Yang-Baxter equation, both the constant equation and the ones with spectral parameters, are denoted using $R$. {We use $\check{R}=PR$ for the analogous braided forms of the Yang-Baxter equation.}
    \item The three Pauli matrices are denoted using $X$, $Y$ and $Z$.
    \item For the Majorana fermions we use both $\gamma$ and $\Gamma$. The indices on the latter are just the physical ones, whereas on the former they include both auxiliary and physical ones.
    \item $\lambda$ and $\mu$ are spectral parameters.
    \item The transfer matrix is denoted using $\tau$. Conserved quantities derived from this operator are denoted with ${\cal U}$ and $\mathbb{Q}$.
    \item The fermionic parity is denoted by ${\cal P}$. While we denote its spin-version as ${\sf P}$. These should not be confused with the usual permutation $P$.
\end{enumerate}

\section{A Majorana fermion Yang-Baxter solution}
\label{sec:majorana-solution}
The $R$-matrix is an operator that offers a possible route to the construction of a quantum integrable system. It satisfies an operator relation known as the {\it Yang-Baxter equation} (YBE) \cite{YangCN1967,BAXTER1972193}. For this reason we will also call the $R$-matrix as a {\it Yang-Baxter operator} (YBO). This equation comes in two forms: with and without spectral parameters. With spectral parameters $u$, $v$, the equation reads
\begin{eqnarray}\label{eq:YBE-spec-nonbraided}
R_{ij}(u-v)R_{ik}(u)R_{jk}(v) = R_{jk}(v)R_{ik}(u)R_{ij}(u-v)~;~u,v\in\mathbb{C}.
\end{eqnarray}
The form of the YBE written in \eqref{eq:YBE-spec-nonbraided} is non-braided as compared to the braided form 
\begin{eqnarray}\label{eq:YBE-spec-braided}
    {\check{R}_{ij}(u-v)\check{R}_{jk}(u)\check{R}_{ij}(v) = \check{R}_{jk}(v)\check{R}_{ij}(u)\check{R}_{jk}(u-v).}
\end{eqnarray} 
The latter has an index structure that resembles the relations satisfied by braid group generators. The non-braided form is more useful in physics as integrability is more apparent from this form. The two forms are related, as if $R$ satisfies the non-braided form, then $PR$ satisfies the braided form and vice versa. Here $P$ is the standard permutation operator satisfying the relations
\begin{eqnarray}\label{eq:perm-relations}
     & P_{ij}P_{jk}P_{ij}=P_{jk}P_{ij}P_{jk}=P_{ik},~P^2=\mathbb{1},~P_{ij}P_{kl} = P_{kl}P_{ij}. & 
\end{eqnarray}
The last relation holds when $j$ and $k$ are not nearest neighbors. It is also known as the far commutativity condition. Such $P$'s generate the permutation group $\mathcal{S}_N$ of order $N$. In our context $N$ will stand for the number of sites of a closed one dimensional lattice.
As an example on $\mathbb{C}^2\otimes\mathbb{C}^2$ we have,
\begin{eqnarray}\label{eq: tensor-perm}
    P=\begin{pmatrix}
    1 & 0 & 0 & 0 \\ 0 & 0 & 1 & 0 \\ 0 & 1 & 0 & 0 \\ 0 & 0 & 0 & 1
\end{pmatrix}=\frac{\mathbb{1}+X\otimes X + Y\otimes Y + Z\otimes Z}{2}.
\end{eqnarray}
The last expression is written in terms of the Pauli matrices, $X$, $Y$ and $Z$, in the basis where $Z$ is diagonal.

Usually the YBE is an operator equation that acts non-trivially on $V_i\otimes V_j\otimes V_k$, with $V$ a local Hilbert space. The notation $R_{ij}$ implies that this operator acts non-trivially on $V_{i}$, $V_{j}$ and trivially, as the identity operator on $V_k$. In this sense the YBE is a local constraint relation. However in this work we use YBO's that are constructed out of Majorana fermions that are inherently non-local. This would seemingly contradict the above statement of the YBE being a local relation. This confusion is resolved when we consider the other form of the YBE, without spectral parameters or otherwise known as the constant YBE. In the non-braided form this equation looks like
\begin{eqnarray}\label{eq:YBE-const-nonbraided}
R_{ij}R_{ik}R_{jk} = R_{jk}R_{ik}R_{ij},
\end{eqnarray}
and the more familiar constant YBE in the braided form reads,
\begin{eqnarray}\label{eq:YBE-const-braided}
{\check{R}_{ij}\check{R}_{jk}\check{R}_{ij} = \check{R}_{jk}\check{R}_{ij}\check{R}_{jk}.}
\end{eqnarray}
The braided $\check{R}$-matrices generate the $N$ strand braid group $\mathcal{B}_N$, the infinite dimensional generalization of the permutation group $\mathcal{S}_N$ that we saw earlier. As before, in many-body systems we can think of $N$ as the number of sites of the one dimensional chain. Solutions to these equations furnish representations of $\mathcal{B}_N$. These braid group representations can be either global or local. By this we mean the following. Consider the constant YBE in the braided form \eqref{eq:YBE-const-braided}. As an equation on $V\otimes V\otimes V$, this can be rewritten as 
\begin{eqnarray}\label{eq:YBE-constant-braided-local}
   {\left(\check{R}\otimes I\right)\left(I\otimes \check{R}\right)\left(\check{R}\otimes I\right) = \left(I\otimes \check{R}\right)\left(\check{R}\otimes I\right)\left(I\otimes \check{R}\right).} 
\end{eqnarray}
This equation is local as solving this equation, which amounts to finding the matrix $R$, is sufficient to generate the full braid group $\mathcal{B}_N$. For instance, finding the $\check{R}$ matrix by solving the equation on say, $V_1\otimes V_2\otimes V_3$ will help us build the other generators by use of the permutation operator. In contrast to this, for non-local representations it is necessary to solve the equations for all the $(N-1)$ generators to define the associated braid group. Examples include the Fibonacci anyon representation \cite{Kauffman_2016,Kauffman_2018} for the latter and the Jones representation {\it via} matrix representations of the Temperley-Lieb algebra \cite{Temperley:1971iq,Kulish_2008} for the former. While this is common knowledge in mathematics, it is not all too familiar in the physics community as most $R$-matrices appearing in quantum integrability are of the local type. 
In this sense the Majorana fermion representations that we will construct and use in this work are non-local representations of the YBE's. We find solutions for all types of YBE's, with and without spectral parameters. This is to be seen as a point of departure from the usual $R$-matrices found in quantum integrable systems.   

Before we write down the Majorana fermion solutions to the YBE we recall basic facts about Majorana fermions and complex fermions that are useful for our purposes. Consider the set of $N$ Majorana fermions $\gamma_j$ satisfying the relations
\begin{eqnarray}\label{eq:Majorana-relations}
   & \gamma_j = \gamma_j^\dag~~\forall~j, & \nonumber \\ 
   & \left\{\gamma_j,\gamma_k\right\} = 2\delta_{jk},~~\forall~j, k\in\{1,2,\cdots, N\}.  &
\end{eqnarray}
These are also interpreted as real fermions as a pair of Majoranas forms complex Dirac fermions through the realization:
\begin{eqnarray}\label{eq:Majorana2complex}
    c_j = \frac{1}{2}\left(\gamma_{2j-1} +\mathrm{i}\gamma_{2j}\right)~;~c_j^\dag = \frac{1}{2}\left(\gamma_{2j-1} -\mathrm{i}\gamma_{2j}\right).
\end{eqnarray}
This mapping makes sense only when there are an even number of Majorana fermions. This is the indeed the most relevant case in this work as we shall soon see. It is now easily seen that the $c$'s along with their adjoints $c^\dag$'s form the {\it canonical anticommutation relations} (CAR) or complex fermion algebra:
\begin{eqnarray}\label{eq:fermions-CAR}
 & \{c_j, c_k\} = \{c_j^\dag , c_k^\dag\} = 0,  & \nonumber \\
 & \{c_j, c_k^\dag\} = \delta_{jk}.    &
\end{eqnarray}
These relations follow from the mapping \eqref{eq:Majorana2complex} and the algebra satisfied by the Majorana fermions \eqref{eq:Majorana-relations}.

Next we require a notion of regularity for the $R$-matrices we construct. This is crucial as regularity for $R$-matrices is a sufficient condition for the existence of local Hamiltonians in the tower of mutually commuting operators that appear from the YBE. As our goal is to find local Hamiltonians, it is important to clarify the existence of regular $R$-matrices that are constructed using Majorana fermions. We will take 
\begin{eqnarray}\label{eq:regularity}
  R(0) = \tilde{P},~~{\check{R}(0) \propto \mathbb{1}}
\end{eqnarray}
where $\tilde{P}$ is a permutation-like operator that can act on the space of Majorana fermion operators. There are several such choices possible as we shall soon see. {It should be noted here that since the $R$-matrix we construct does not reduce to the usual permutation operator $P$, we consider these $R$-matrices to be non-regular.}

\begin{remark}
    {It should be noted that transfer matrices of the two dimensional classical Ising model can also be constructed using Temperley-Lieb (TL) generators \cite{Koo_1994,Potts_PPMartin}. The TL generators in this case are also realized using Majorana fermions [See Sec. 3 of \cite{Koo_1994}]. However these operators satisfy the braided form of the YBE and do not commute with the TFIM Hamiltonian as explained in Sec. \ref{sec:Introduction}. }
    
\end{remark}

With these ingredients in place we can write down algebraic solutions for the YBE in both the constant and spectral parameter dependent form. We start with a solution of the former. Consider the non-local $R$-matrix constructed out of Majoranas\footnote{We call $\tilde{P}$ as $P^-$ as we can define another permutation-like operator $$ P^+_{jk} = \frac{\gamma_j+\gamma_k}{\sqrt{2}}. $$ Its definition and consequences to integrable models are discussed in Appendix \ref{app:P+}.}
\begin{eqnarray}\label{eq:P-Majorana}
    \tilde{P}_{jk} = P^-_{jk} = \frac{\gamma_j-\gamma_k}{\sqrt{2}}.
\end{eqnarray}
The reason for denoting this $R$-matrix as $P^-$ becomes clear from the permutation-like properties of this operator. For instance the inverse is itself
\begin{eqnarray}\label{eq:P-inverse-Majorana}
   (P^-_{jk})^{-1}=P^-_{jk} = \frac{\gamma_j-\gamma_k}{\sqrt{2}}. 
\end{eqnarray}
It satisfies both the braided and non-braided forms of the constant YBE as we have
$$ P^-_{ij}P^-_{jk}P^-_{ij}=P^-_{jk}P^-_{ij}P^-_{jk} = P^-_{ik},  $$
as expected from a permutation operator. However it satisfies a far-anticommutativity relation
\begin{eqnarray}
    P^-_{ij}P^-_{kl} = -P^-_{kl}P^-_{ij}, 
\end{eqnarray}
instead of the far-commutativity relation. This distinguishes it from the usual permutation operator. A consequence of this is seen in the way this permutation interchanges other Majorana fermion operators. The action is by conjugation
\begin{eqnarray}\label{eq:ptilde-action-majoranas}
    &P^-_{jk}\gamma_jP^-_{jk} = -\gamma_k~;~ P^-_{jk}\gamma_k P^-_{jk} = -\gamma_j, & \nonumber \\
    &  P^-_{jk}\gamma_l P^-_{jk} = -\gamma_l,~~~\textrm{when}~l\neq j,k.   &
\end{eqnarray}
Thus, unlike the usual permutation operator the Majorana permutation operator interchanges indices up to an overall negative sign. We would expect this to generalize to Majorana parafermions which would  interchange indices by a similar exchange operator, but up to an overall phase that is a root of unity. 

Next we consider a spectral parameter dependent solution. The Majorana fermion operator 
\begin{eqnarray}\label{eq:R-spec-Majorana-braided}
    {\check{R}_{jk} = \mathbb{1} + \mathrm{i}\tan{\left(\lambda\right)}~\gamma_j\gamma_k,~~~\mathrm{i}=\sqrt{-1},~\lambda\in\mathbb{C},}
\end{eqnarray}
with the spectral parameter $\lambda$ satisfies the spectral parameter dependent YBE \eqref{eq:YBE-spec-braided}. This is a braided $\check{R}$-matrix.
Its inverse is given by 
$${\check{R}_{jk}^{-1} = \frac{\mathbb{1} - \mathrm{i}\tan{\left(\lambda\right)}~\gamma_j\gamma_k}{1-\tan^2({\lambda})}.} $$

\begin{remark}\label{rem:generalized-regular-R}
  This solution can be seen as the Baxterized\footnote{{\it Baxterization} is a generic method to produce spectral parameter dependent $R$-matrices from braid group generators [constant YBO's] \cite{Jones1990}.} version of the braid group generator 
\begin{eqnarray}\label{eq:b-Majorana}
    \rho_{jk}=\frac{\mathbb{1}+\gamma_j\gamma_k}{\sqrt{2}}.
\end{eqnarray}
This is the Majorana braid operator that appears in the context of $p$-wave superconductor \cite{Ivanov2001} and also in fermionic quantum computing \cite{Bravyi_2002}. Unlike the Majorana permutation operator in \eqref{eq:P-Majorana}, this operator does not square to the identity operator. Instead we have 
$$ \rho^{-1}_{jk}=\frac{\mathbb{1}-\gamma_j\gamma_k}{\sqrt{2}}~;~\rho^8=\mathbb{1}.$$
This operator satisfies the far-commutativity relation as expected off a braid group generator. However it exchanges Majorana fermions in a different way:
\begin{eqnarray}\label{eq:bMajorana-action-majoranas}
    &\rho_{jk}\gamma_j\rho_{jk}^{-1} = -\gamma_k~;~ \rho_{jk}\gamma_k \rho_{jk}^{-1} = \gamma_j, & \nonumber \\
    &  \rho_{jk}\gamma_l \rho_{jk}^{-1} = \gamma_l,~~~\textrm{when}~l\neq j,k.   &
\end{eqnarray}
This exchange property is not usual for generic braid group representations. But due to this feature, we can use the Majorana braid generator in place of the permutation operator in constructing integrable models. This essentially implies that we can broaden the notion of a regular $R$-matrix by demanding that
$$ R_{jk}(0) = \rho_{jk}.  $$ 
We leave the exploration of its consequences to the future.
\end{remark}

To obtain the non-braided form that satisfies the physically relevant non-braided YBE \eqref{eq:YBE-spec-nonbraided}, we multiply this $R$-matrix with the Majorana fermion permutation operator $P^-$,
\begin{eqnarray}\label{eq:R-spec-Majorana-nonbraided}
    R_{jk}(\lambda) = \frac{\gamma_j-\gamma_k}{\sqrt{2}}\left[\mathbb{1} + \mathrm{i}\tan{\left(\lambda\right)}~\gamma_j\gamma_k\right] = \frac{1+\mathrm{i}~\tan({\lambda})}{\sqrt{2}}\left(\gamma_j + f_\lambda~\gamma_k\right),
\end{eqnarray}
with $f_\lambda = -\left[\frac{1-\mathrm{i}~\tan({\lambda})}{1+\mathrm{i}~\tan({\lambda})} \right].$ This $R$-matrix is regular in the sense defined earlier as $R(0) = P^-$, {but it is non-regular in the traditional definition where the $R$-matrix has to reduce to the usual permutation operator. To reflect this we will henceforth call our $R$-matrices {\it `regular'} in quotes to distinguish it from the usual notion of regularity.} The factor $\frac{\left[1+\mathrm{i}~\tan({\lambda})\right]}{\sqrt{2}}$ is an overall scaling function, and so does not alter the properties of the constructed $R$-matrix. We will ignore it in the rest of the computations. We will use both these forms  
\begin{eqnarray}\label{eq:R-matrix-form1-form2}
    R_{jk}(\lambda) = \begin{cases}
        \frac{\gamma_j-\gamma_k}{\sqrt{2}}\left[\mathbb{1} + \mathrm{i}\tan{\left(\lambda\right)}~\gamma_j\gamma_k\right]&\textrm{Form I}, \\
       \gamma_j + f_\lambda~\gamma_k~;~f_\lambda = -\left[\frac{1-\mathrm{i}~\tan({\lambda})}{1+\mathrm{i}~\tan({\lambda})} \right]&\textrm{Form II},
    \end{cases}
\end{eqnarray}
for the $R$-matrix as demanded by the situation. 

\section{Commutativity of transfer matrices}
\label{sec:commutativity-transfer}
Now that we have a `regular', spectral parameter dependent $R$-matrix constructed out of Majorana fermions, we are in a position to find the mutually commuting set of operators from a transfer matrix. The first step for this is to write down the monodromy matrix $T$ in terms of the Lax operators $L$\footnote{We use the following convention: 
\begin{eqnarray}
    \prod_{j=N}^1{\cal O}_j={\cal O}_N\cdots{\cal O}_1.
\end{eqnarray}}:
\begin{eqnarray}\label{eq:monodromy}
    T_a(\lambda) = \prod\limits_{j=N}^1~L_{aj}(\lambda) = \prod\limits_{j=N}^1~R_{aj}(\lambda),\quad {T_b(\lambda) = \prod\limits_{j=N}^1~L_{bj}(\lambda) = \prod\limits_{j=N}^1~R_{bj}(\lambda).}
\end{eqnarray}
Here $a,b$ denotes the indices for the auxiliary spaces that will eventually get traced over. The $N$ physical space indices are denoted using $j$. The Lax operators themselves satisfy the so called $RLL$ relation
\begin{eqnarray}\label{eq:RLL}
    R_{ab}(\lambda-\mu)L_{aj}(\lambda)L_{bj}(\mu) = L_{bj}(\mu)L_{aj}(\lambda)R_{ab}(\lambda-\mu).
\end{eqnarray}
A possible solution for the Lax operator is the $R$-matrix itself. This is called the fundamental representation and is the one used to construct the monodromy matrix in \eqref{eq:monodromy}. For the Majorana solutions this $R$-matrix can take the forms I or II in \eqref{eq:R-matrix-form1-form2}.
Using this expression we will first verify the that the monodromy matrix satisfies the spectral parameter dependent YBE in \eqref{eq:YBE-spec-nonbraided},
\begin{eqnarray}\label{eq:RTT}
    R_{ab}(\lambda-\mu)T_a(\lambda)T_b(\mu) = T_b(\mu)T_a(\lambda)R_{ab}(\lambda-\mu).
\end{eqnarray}
This is also known as the $RTT$ relation in the literature \cite{Korepin1993QuantumIS,slavnov2019algebraicbetheansatz}. For local $R$-matrices this is a rather straightforward exercise \cite{slavnov2019algebraicbetheansatz}, but requires a proof for the non-local $R$-matrices considered here. Such a proof goes as follows:
\begin{eqnarray}\label{eq:proof-RTT}
    R_{ab}(\lambda-\mu)T_a(\lambda)T_b(\mu) & = & R_{ab}\left[\prod\limits_{j=N}^1\left(\gamma_a+f_\lambda~\gamma_{j} \right)\right]\left[\prod\limits_{k=N}^1\left(\gamma_b+f_\mu~\gamma_{k} \right)\right] \nonumber \\
    & = & \left(-1\right)^{\frac{N(N-1)}{2}}R_{ab}\left[\prod\limits_{j=N}^1~\left(\gamma_a+f_\lambda~\gamma_{j} \right) \left(\gamma_b+f_\mu~\gamma_{j} \right)\right] \nonumber \\ & = & \left(-1\right)^{\frac{N(N-1)}{2}} \left[\prod\limits_{j=N}^1~\left(\gamma_b+f_\mu~\gamma_{j} \right) \left(\gamma_a+f_\lambda~\gamma_{j} \right)\right]R_{ab} \nonumber \\
    & = & \left(-1\right)^{N(N-1)}\left[\prod\limits_{k=N}^1\left(\gamma_b+f_\mu~\gamma_{k} \right)\right]\left[\prod\limits_{j=N}^1\left(\gamma_a+f_\lambda~\gamma_{j} \right)\right]R_{ab} \nonumber \\
    & = & T_b(\mu)T_a(\lambda) R_{ab}(\lambda-\mu).
\end{eqnarray}
We have suppressed the $(\lambda-\mu)$ dependence of the $R$-matrix on most lines of the right hand side for brevity. To go from the second to the third line we have repeatedly used the fact that the Majorana $R$-matrix \eqref{eq:R-matrix-form1-form2} satisfies the YBE in \eqref{eq:YBE-spec-nonbraided}. Also, $N(N-1)$ is even for both odd and even $N$, making the monodromy matrix in \eqref{eq:monodromy} satisfy the $RTT$ relation \eqref{eq:RTT} for all $N$.

Let us assume that we have even $2N$ number of Majorana modes. We shall come back to the case of odd Majorana modes later. First we notice that the operators $T_{a(b)}(x)$ can be expanded as
\begin{eqnarray}\label{eq:T-odd-even}
    T_{a(b)}(x)=\tau(x)+\gamma_{a(b)}\sigma(x)~;~x\in\{\lambda, \mu\}.
\end{eqnarray}
It is then evident that $\tau(x)$ is made up of operators which have an even number of Majorana modes $\gamma_m\neq\gamma_{a},\gamma_b$. On the other hand, $\sigma(x)$ consists of those terms which have an odd number of $\gamma_m\neq\gamma_{a},\gamma_b$. Then we obtain the following relations:
\begin{eqnarray}
    \gamma_{a(b)}\tau(x)=\tau(x)\gamma_{a(b)},\quad\gamma_{a(b)}\sigma(x)=-\sigma(x)\gamma_{a(b)}.
\end{eqnarray}
Next, it is easy to check that $R_{ab}(x)^{-1}=\left[{1}/({1+f_x^2})\right]R_{ab}(x)$, which allows us to rewrite the $RTT$ relation as
\begin{eqnarray}\label{eq:RTT-proof-1}
    R_{ab}(\lambda-\mu)T_a(\lambda)T_b(\mu)R_{ab}(\lambda-\mu)=(1+f_{\lambda-\mu}^2)T_b(\mu)T_a(\lambda).
\end{eqnarray}
We further can expand the left hand side as
\begin{eqnarray}
    &&R_{ab}(\lambda-\mu)T_a(\lambda)T_b(\mu)R_{ab}(\lambda-\mu)\nonumber\\
    &=&(\gamma_a+f_{\lambda-\mu}\gamma_b)[\tau(\lambda)+\gamma_a\sigma(\lambda)][\tau(\mu)+\gamma_b\sigma(\mu)](\gamma_a+f_{\lambda-\mu}\gamma_b)\nonumber\\
    &=&\gamma_a{{\cal M}(\lambda,\mu)}\gamma_a+f_{\lambda-\mu}\left[\gamma_a{{\cal M}(\lambda,\mu)}\gamma_b+\gamma_b{{\cal M}(\lambda,\mu)}\gamma_a\right]+f_{\lambda-\mu}^2\gamma_b{{\cal M}(\lambda,\mu)}\gamma_b,
\end{eqnarray}
with 
\begin{eqnarray}
    {{\cal M}(\lambda,\mu)}&=&[\tau(\lambda)+\gamma_a\sigma(\lambda)][\tau(\mu)+\gamma_b\sigma(\mu)]\nonumber\\    &=&\tau(\lambda)\tau(\mu)+\tau(\lambda)\gamma_b\sigma(\mu)+\gamma_a\sigma(\lambda)\tau(\mu)+\gamma_a\sigma(\lambda)\gamma_b\sigma(\mu).
\end{eqnarray}

{Till now, our entire discussion has been purely algebraic. One now should perform partial traces over the auxiliary indices to arrive at the commutativity of the transfer matrices. At this point we anticipate issues arising from the partial traces of the Majorana fermion operators. However all impending ambiguities are laid to rest by invoking the qubit/spin $\frac{1}{2}$ realization of the Majorana fermions using the {\it Jordan-Wigner transformation} (JW). This requires a convention for the ordering of the indices. In what follows, we shall consider the following arrangement:
\begin{eqnarray}\label{eq:JW-transform}
    & \gamma_a = X_a,~~\gamma_b=Z_a\left[\prod\limits_{j=1}^{N}Z_j\right]X_b, & \nonumber \\
    & \gamma_{2k-1} = Z_a\left[\prod\limits_{j=1}^{k-1}Z_j\right]X_k,~~ \gamma_{2k} = Z_a\left[\prod\limits_{j=1}^{k-1}Z_j\right]Y_k,\quad k\in\{1,\cdots, N\}.&
\end{eqnarray}
Notably, the Majorana modes with consecutive odd $(2k-1)$ and even $(2k)$ sites are mapped to the same index $k$ in the spin representation. This helps us halve the total number of sites from $2N$ to $N$.  It is useful for later purposes to introduce the notation 
\begin{eqnarray}\label{eq:Gamma-Majorana}
    &\Gamma_{2k-1} = \left[\prod\limits_{j=1}^{k-1}Z_j\right]X_k,~~\Gamma_{2k} = \left[\prod\limits_{j=1}^{k-1}Z_j\right]Y_k,\quad k\in\{1,\cdots, 2N\},&
\end{eqnarray}
to denote Majoranas devoid of the auxiliary indices $a,b$. Specifically, we have the following identification:
\begin{eqnarray}
    \gamma_j=Z_a\Gamma_j,\quad j\in\{1,\cdots, 2N\}.
\end{eqnarray}
The new $\Gamma$'s satisfy all the algebraic properties of the Majorana modes, as one can verify easily. Since we are working with even number of Majorana modes and $\gamma_j\gamma_k=\Gamma_j\Gamma_k,~\forall j,k\in\{1,\cdots,2N\}$, the operator $\tau(\lambda)$ acts trivially on the auxiliary indices $a$ and $b$. This allows us to introduce the transfer matrix $\tau(\lambda)$ as
\begin{eqnarray}\label{eq:transfer-matrix}
    \tau(\lambda)={\rm tr}_{a}[T_{a}(\lambda)]={\rm tr}_b[T_b(\lambda)].
\end{eqnarray}}

As we trace out the auxiliary indices, the non-zero terms do not contain either of $\gamma_a$ or $\gamma_b$. Therefore, the terms in ${\cal M}$ which are linear in either 
$\gamma_a$ or $\gamma_b$ do not contribute. Consequently, we have
{\begin{eqnarray}
    &{\rm tr}_{a,b} [\gamma_a{{\cal M}(\lambda,\mu)}\gamma_a]={\rm tr}_{a,b} [\gamma_b{{\cal M}(\lambda,\mu)}\gamma_b]=\tau(\lambda)\tau(\mu),&\nonumber\\
    &{\rm tr}_{a,b} [\gamma_a{{\cal M}(\lambda,\mu)}\gamma_b+\gamma_b{{\cal M}(\lambda,\mu)}\gamma_a]=0.&
\end{eqnarray}}
Summing these we obtain
\begin{eqnarray}
    {\rm tr}_{a,b}\left[R_{ab}(\lambda-\mu)T_a(\lambda)T_b(\mu)R_{ab}(\lambda-\mu)\right]=(1+f_{\lambda-\mu}^2)\tau(\lambda)\tau(\mu).
\end{eqnarray}
Now substituting this along with ${\rm tr}_{a,b}[T_b(\mu)T_a(\lambda)]=\tau(\mu)\tau(\lambda)$ back into \eqref{eq:RTT-proof-1} yields the commutation relation
\begin{eqnarray}\label{eq:TMcom}
    [\tau(\lambda),\tau(\mu)]=0.
\end{eqnarray}
This shows that the Majorana transfer matrices commute for even number of Majorana modes. It should be noted that we have avoided using the cyclic property under the partial trace in arriving at this result. This property is used to show the commutativity relation \eqref{eq:transfer-matrix} for local $R$-matrices. 

\begin{remark}
    {The above proof for the commutativity of transfer matrices only requires that the Majorana fermions [or an odd number of them] are traceless : $$ \tr\left[ \gamma \right] = \tr\left[ \Gamma \right] =0.$$
    This follows once we realize the Majorana fermions as a string of Clifford algebra generators, generalizing the JW transforms in \eqref{eq:JW-transform}. Clifford algebra generators are traceless in finite dimensions, thus making the above proof representation independent.}
\end{remark}

The above arguments fail when one considers an odd number of Majorana modes. It turns out that, The transfer matrix $\tau(\lambda)$ itself is well defined only when there are even number of Majorana modes. As we shall now show, for odd number of modes, the transfer matrix at $\lambda=0$ vanishes identically. To this end, let us consider $\tau(0)={\rm tr}_a[T_a(0)]$ as 
\begin{eqnarray}
     &&{\rm tr}_a\left[(\gamma_a-\gamma_{\cal N})\cdots(\gamma_a-\gamma_1)\right]=(-1)^{{\cal N}-1}{\rm tr}_a\left[(\gamma_a-\gamma_1)(\gamma_1-\gamma_{\cal N})\cdots(\gamma_1-\gamma_2)\right],
\end{eqnarray}
where, ${\cal N}$ is $2N$ for even and is $2N+1$ for odd number of Majorana modes.
Using the spin-realization \eqref{eq:JW-transform} of the Majorana modes, the above trace can be expressed as
\begin{eqnarray}\label{eq:even-odd TM}
    &&{\tr}_a\left[X_aZ_a^{{\cal N}-1}(\Gamma_1-\Gamma_{\cal N})\cdots(\Gamma_1-\Gamma_2)-Z_a^{\cal N}\Gamma_1(\Gamma_1-\Gamma_{\cal N})\cdots(\Gamma_1-\Gamma_2)\right].
\end{eqnarray}
Now, ${\rm tr}_a[X_aZ_a^{{\cal N}-1}]=0,~\forall {\cal N}$, and ${\rm tr}_a[Z_a^{\cal N}]=0$ whenever we have odd ${\cal N}$, rendering $\tau(0)=0$ for odd number of Majorana modes. This shows that the construction of an integrable Majorana system inherently demands an even number of Majorana modes.\footnote{{In fact, we can find commuting transfer matrices for systems with odd Majorana modes as well if we relax the requirement of extracting the transfer matrix by naively executing the partial trace on the monodromy matrix. More details can be found in the appendix App.\ref{app:oddMaj}.}} Furthermore, we have ${\rm tr}_a[Z_a^{\cal N}]=2$ for even ${\cal N}$, yielding
\begin{eqnarray}\label{eq:transfer-matrix-lambda0}
    \tau(0)=2\Gamma_1(\Gamma_1-\Gamma_{\cal N})\cdots(\Gamma_1-\Gamma_2).
\end{eqnarray}
The invertibility of $\tau(0)$ plays a crucial role in the construction of the Hamiltonian. Furthermore, as is the case for the usual  integrable system (defined on factorized Hilbert space) constructed from \textit{regular} $R$-matrix, the transfer matrix $\tau(0)$ behaves very much like the conserved translation operator of the system.

{
\begin{remark}
    The operator $\tau(0)$ helps us anticipate the nature of the integrable Hamiltonian. Let us observe the action of it on the boundary Majorana modes $\Gamma_1,\Gamma_{2N}$. One easily finds out that
    \begin{eqnarray}
        \tau(0)~\Gamma_{1}~\tau(0)^{-1}=\Gamma_2,\quad \tau(0)~\Gamma_{2N}~\tau(0)^{-1}=-\Gamma_1.
    \end{eqnarray}
    This immediately implies that the integrable Hamiltonian, which by the very construction should commute with $\tau(0)$, cannot be translation invariant. Rather, it strongly suggests the presence of an anti-periodic boundary term in the Hamiltonian. In what follows, we shall derive the Hamiltonian and show that it breaks the translation invariance explicitly.
\end{remark}
}

\section{Integrable local Hamiltonian}
\label{sec:critical-Ising}
With the integrability well established for even number of sites, we can now systematically construct the Hamiltonian and commuting tower of conserved quantities from the transfer matrix \eqref{eq:transfer-matrix}. We will do this at the point $\lambda=0$, where the Majorana fermion $R$-matrix becomes `regular' in the sense defined earlier. Formally expanding the transfer matrix as $\log \tau(\lambda)=\mathbb{Q}_1+\lambda\mathbb{Q}_2+\lambda^2\mathbb{Q}_3+\cdots$, we obtain, from Eqn. \eqref{eq:TMcom},
\begin{eqnarray}\label{eq:consq}
    \left[{\mathbb Q}_m,{\mathbb Q}_n\right]=0,\quad \forall m,n.
\end{eqnarray}
These mutually commuting conserved charges generate the Abelian algebra ${\cal Q}$. Every element of ${\cal Q}$ also commutes with the Hamiltonian and thus remains conserved. The integrable Hamiltonian typically is associated with the charge $\mathbb{Q}_2$, which can be expressed as
\begin{eqnarray}
    H={\tau}(0)^{-1}\frac{{\rm d}{\tau}(\lambda)}{{\rm d}\lambda}\Bigg|_{\lambda=0}. 
\end{eqnarray}
It is also called trace identities.\\
This computation yields a local Hamiltonian for the `regular', non-local $R$-matrices found here just as in the case of regular, local $R$-matrices. To see this we first compute the derivative 
\begin{eqnarray}
    -{\rm i}\frac{{\rm d}{\tau}(\lambda)}{{\rm d}\lambda}\Bigg|_{\lambda=0}&=&{\rm tr}_a\left[(\gamma_a-\gamma_{2N})\cdots(\gamma_a-\gamma_1)\gamma_a\gamma_1\right]\nonumber\\
    &&\quad\quad\quad+\sum_{j=2}^{2N} {\rm tr}_a\left[(\gamma_a-\gamma_{2N})\cdots(\gamma_a-\gamma_j)\gamma_a\gamma_j(\gamma_a-\gamma_{j-1})\cdots(\gamma_a-\gamma_1)\right]\nonumber\\
    &=&-{\rm tr}_a\left[(\gamma_a-\gamma_1)(\gamma_1-\gamma_{2N})\cdots(\gamma_1-\gamma_2)\gamma_a\gamma_1\right]\nonumber\\
    &&\quad\quad\quad-\sum_{j=2}^{2N} {\rm tr}_a\left[(\gamma_a-\gamma_1)(\gamma_1-\gamma_{2N})\cdots(\gamma_1-\gamma_2)\gamma_{j-1}\gamma_j\right].
\end{eqnarray}
We will now use the JW transform of \eqref{eq:JW-transform} to evaluate these expressions.
\begin{eqnarray}
    -{\rm i}\frac{{\rm d}{\tau}(\lambda)}{{\rm d}\lambda}\Bigg|_{\lambda=0}&=& -{\rm tr}_a\left[(X_a-Z_a\Gamma_1)Z_a(\Gamma_1-\Gamma_{2N})\cdots Z_a(\Gamma_1-\Gamma_2)X_aZ_a\Gamma_1\right]\nonumber\\
    &&-\sum_{j=2}^{2N} {\rm tr}_a\left[(X_a-Z_a\Gamma_1)Z_a(\Gamma_1-\Gamma_{2N})\cdots Z_a(\Gamma_1-\Gamma_2)Z_a\Gamma_{j-1}Z_a\Gamma_j\right]\nonumber\\
     &=&-{\rm tr}_a\left[X_aZ_a(\Gamma_1-\Gamma_{2N})\cdots Z_a(\Gamma_1-\Gamma_2)(-Z_aX_a)\Gamma_1\right]\nonumber \\
      &&-\sum_{j=2}^{2N} {\rm tr}_a\left[(-Z_a\Gamma_1)Z_a(\Gamma_1-\Gamma_{2N})\cdots Z_a(\Gamma_1-\Gamma_2)Z_a\Gamma_{j-1}Z_a\Gamma_j\right].
\end{eqnarray}
Note that isolating the auxiliary index $a$ helps us evaluate the partial traces easily. Doing so gives us
\begin{eqnarray}
    -{\rm i}\frac{{\rm d}{\tau}(\lambda)}{{\rm d}\lambda}\Bigg|_{\lambda=0}&=&2(\Gamma_1-\Gamma_{2N})\cdots (\Gamma_1-\Gamma_2)\Gamma_1+2\sum_{j=2}^{2N} \Gamma_1(\Gamma_1-\Gamma_{2N})\cdots (\Gamma_1-\Gamma_2)\Gamma_{j-1}\Gamma_j\nonumber\\
    &=&2\Gamma_1\Gamma_1(\Gamma_1-\Gamma_{2N})\cdots (\Gamma_1-\Gamma_2)\Gamma_1+{\tau}(0)\sum_{j=2}^{2N} \Gamma_{j-1}\Gamma_j\nonumber\\
     &=&-2\Gamma_1(\Gamma_1-\Gamma_{2N})\cdots (\Gamma_1-\Gamma_2)\Gamma_{2N}\Gamma_1+{\tau}(0)\sum_{j=2}^{2N} \Gamma_{j-1}\Gamma_j\nonumber\\
     &=&{\tau}(0)\left(-\Gamma_{2N}\Gamma_1+\sum_{j=2}^{2N}\Gamma_{j-1}\Gamma_j\right).
\end{eqnarray}
Multiplying this with the inverse of $\tau(0)$ in \eqref{eq:transfer-matrix-lambda0} yields the closed chain Hamiltonian with anti-periodic boundary conditions written in terms of Majorana fermions,
\begin{eqnarray}\label{eq:kitaevH-antiperiodic}
   H ={\rm i}\sum_{j=1}^{2N-1}\Gamma_{j}\Gamma_{j+1}-{\rm i}\Gamma_{2N}\Gamma_1.
\end{eqnarray}
Interestingly, it also describes a Majorana chain with fermionic parity defect in the link $(2N,1)$ (see Eqn. $69$, of \cite{seiberg2024majorana}). In the thermodynamic limit, $N\to\infty$, the boundary term can be ignored and the Hamiltonian above then rightly describes the $1D$ Kitaev chain at criticality \cite{Kitaev_2001}. In deriving this Hamiltonian we have not used the cyclicity under the partial trace at any stage of the computation. We have merely used the index exchanging properties of the Majorana permutation operator $(\gamma_j-\gamma_k)$,  \eqref{eq:ptilde-action-majoranas} to simplify the expressions and facilitate the evaluation of the partial traces. The entire method can be seen as a substitute for the trace identities \cite{Korepin1993QuantumIS} which is traditionally used to evaluate the local Hamiltonian from transfer matrices built out of local, regular $R$-matrices. 

\subsection{The critical Ising Hamiltonian}
\label{subsec:Ising-H}
We will now make the explicit connection of the Hamiltonian in \eqref{eq:kitaevH-antiperiodic} to the TFIM. To begin with, let us briefly talk about the different phases of the TFIM, described by the Hamiltonian \eqref{eq:Ising-H}. When the strength of the transverse magnetic field $h$ is smaller than the interaction strength $J$, i.e. when $J>h$, the system is ferromagnetic and the ground state is two-fold degenerate where the spins tend to align along the $x$-direction as the $X_jX_{j+1}$ interaction term dominates over the magnetic field. Furthermore, the $\mathbb{Z}_2$ symmetry connecting the two ground states is spontaneously broken. On the other hand, when the strength of the magnetic field $h$ overcomes the interaction strength $J$, i.e. in the region $h>J$, there is a unique ground state. The above two phases are connected by the KW duality. When $h=J$, the system undergoes a quantum phase transition and becomes critical. This critical point belongs to the $(1+1)$-D Ising universality class and is described by a \textit{conformal field theory} (CFT) with central charge $c=1/2$. At this point, the KW duality becomes a symmetry of the system. 

In what follows, we will show that the integrable Majorana Hamiltonian \eqref{eq:kitaevH-antiperiodic}, is nothing but the critical quantum Ising model, up to an overall scale factor. The Majorana Hamiltonian can also be expressed as
\begin{eqnarray}
    H={\rm i}\sum_{j=1}^N\Gamma_{2j-1}\Gamma_{2j}+{\rm i}\sum_{j=1}^{N-1}\Gamma_{2j}\Gamma_{2j+1}-{\rm i}\Gamma_{2N}\Gamma_1.
\end{eqnarray}
Using the JW transformation \eqref{eq:Gamma-Majorana}, we arrive at the spin-Hamiltonian
\begin{eqnarray}\label{eq:critical-TFIM-finite}
    H=-\sum_{j=1}^{N}Z_j-\sum_{j=1}^{N-1}X_jX_{j+1}-{\sf P}X_NX_1,\quad {\sf P}=Z_1\cdots Z_N,
\end{eqnarray}
with the $(N+1)$th site being identified with the first site. The operator ${\sf P}$ is conserved since $[H,{\sf P}]=0$ and satisfies ${\sf P}^2=\mathbb{1}$. Therefore, when written in terms of the spin operators, the Hamiltonian acquires the non-local term ${\sf P}X_NX_1$. Consequently, the resulting Hamiltonian is not that of the critical TFIM, but differs from it by $\Delta H=H-H_{\rm TFIM}=(1-{\sf P})X_NX_1$. 

However, phenomena such as quantum phase transition and criticality make sense only when the thermodynamic limit, $N\to\infty$, is considered. In this scenario, the corrections appearing due to $\Delta H$ become negligible and hence can be ignored\footnote{An operator becomes irrelevant in the thermodynamic limit, if the ratio of the expectation value of that operator in any quantum state to the system size becomes zero in this limit. Since, $\Delta H^3=4\Delta H$, the limit $\lim_{N\to\infty}(1/N)\langle \Psi|\Delta H|\Psi\rangle$ vanishes identically for any $\Psi$.}. In other words, the Hamiltonian $H$ approaches to $H_{\rm TFIM}$ in the thermodynamic limit. This essentially implies that starting from a well-defined $R$-matrix, we can obtain the critical TFIM, thus making it Yang-Baxter integrable. 
\subsection{A twisted translation operator}
Let us now express $\tau(0)$, which is conserved by the construction, in a somewhat different form. It is easy to check that, up to a factor of $2$, we have
\begin{eqnarray}
    \tau(0)&=&\Gamma_1(\Gamma_1-\Gamma_{2N})\cdots(\Gamma_1-\Gamma_2)\nonumber\\
    &=& (-1)^{2N-1}(\Gamma_1-\Gamma_{2N})\cdots(\Gamma_1-\Gamma_2)\Gamma_{2N}\nonumber\\
    &=&-(-1)^{(N-1)(2N-1)}(\Gamma_1-\Gamma_2)\cdots(\Gamma_{2N-1}-\Gamma_{2N})\Gamma_{2N}\nonumber\\
    &=&(-1)^N(\Gamma_1-\Gamma_2)\cdots(\Gamma_{2N-1}-\Gamma_{2N})\Gamma_{2N}.
\end{eqnarray}
We can now verify that, $\tau(0)^\dagger\tau(0)=2^{2N-1}$, which allows us to define the ``twisted translation operator" ${{\cal U}}$ as
\begin{eqnarray}\label{eq:Majorana-translation-unitary}
    {{\cal U}}=\frac{(-1)^N}{2^{N-\frac{1}{2}}}(\Gamma_1-\Gamma_2)\cdots(\Gamma_{2N-1}-\Gamma_{2N})\Gamma_{2N},\quad {{\cal U}}^{-1}={{\cal U}}^{\dagger}.
\end{eqnarray}
This is nothing but the transfer matrix $\tau(0)$, scaled appropriately to make it unitary, ${\cal U}=\left(1\big/{2^{N-{1}/{2}}}\right){\tau(0)}$. The action of this operator on the Majorana modes can be found to be
\begin{eqnarray}
    {{\cal U}}\Gamma_j{{\cal U}}^{-1}=
    \begin{cases}
        \Gamma_{j+1}\quad\text{for }j=1,\cdots,2N-1\\
    -\Gamma_1\quad\,\text{for }j=2N.
    \end{cases}
\end{eqnarray}
Defining the fermionic parity 
\begin{eqnarray}\label{eq:fermpar}
    {\cal P}:=\Gamma_1\cdots\Gamma_{2N},
\end{eqnarray}
one can verify that the following relations hold:
\begin{eqnarray}
    {\cal U}^{2N}=(-1)^{\frac{N(N+1)}{2}}{\cal P},\quad {\cal P}={\rm i}^N{\sf P}.
\end{eqnarray}
This essentially establishes that the parity ${\sf P}$ is an element of the conserved algebra ${\cal Q}$ and hence commutes with every other element of ${\cal Q}$.

This must be compared to the typical translation operator found, for instance, in the Heisenberg $XXX$-spin chain. In this case, the transfer matrix, at a suitable value of the spectral parameter, yields the translation operator which respects the periodicity of the system. To be explicit, the translation operator $U$ for the $XXX$-spin chain is derived as \cite{faddeev1996algebraic}
\begin{eqnarray}
    U=P_{1,2}P_{2,3}\cdots P_{2N-1,2N},\quad U^{-1}=U^{\dagger},\quad UX_jU^{-1}=X_{j+1},\quad U^{2N}=\mathbb{1},
\end{eqnarray}
 where the operators $P_{j,k}$'s are the usual permutation operators acting on the indices $j$ and $k$ (see Eqn. \eqref{eq: tensor-perm}), with $j,k=1,\cdots, 2N$ and $2N+1$-th site being identified with the first site itself. Therefore, the action of ${{\cal U}}$ differs from that of the usual translation operator on the $2N$-th Majorana mode by a negative sign. 

Using the JW transformation \eqref{eq:Gamma-Majorana}, we can now go to the spin-basis and express the action of ${\cal U}$ on the spin degrees of freedom. It is straightforward to get the action of ${\cal U}$ on the local terms $Z_j=-{\rm i}\Gamma_{2j-1}\Gamma_{2j}$ and $X_jX_{j+1}=-{\rm i}\Gamma_{2j}\Gamma_{2j+1}$, which we summarize in Table \ref{tab: Majorana-Translation}.
\begin{table}[!h]
    \centering
    \begin{tabular}{c c}
        \hline
        \hline
      $\boldsymbol{{\cal O}}$  &  $\boldsymbol{{\cal U}{\cal O}{\cal U}^{-1}}$\\
        \hline
        $Z_j$ & $X_jX_{j+1}$ \\
        $Z_N$ & ${\sf P}X_NX_1$ \\
        $X_jX_{j+1}$ & $Z_{j+1}$ \\
        $X_NX_1$ & ${\sf P}Z_1$\\
        \hline
        \hline
    \end{tabular}
       \caption{The action of the translation operator ${\cal U}$ on different local terms. The index $j$ runs from $1$ to $N-1$.\label{tab: Majorana-Translation}} 
\end{table}
One can check that, ${\cal U}$ leaves the Hamiltonian invariant, i.e. ${\cal U}H{\cal U}^{-1}=H$. To obtain an explicit realization of ${\cal U}$ in the spin basis, we first rearrange ${\cal U}$ in the following way:
\begin{eqnarray}\label{eq:Majorana-translation-KW}
    {{\cal U}}&=&\frac{(-1)^N}{2^{N-\frac{1}{2}}}(\Gamma_1-\Gamma_2)\cdots(\Gamma_{2N-1}-\Gamma_{2N})\Gamma_{2N}\nonumber\\
    &=&\frac{(-1)^N}{2^{N-\frac{1}{2}}}(1+\Gamma_1\Gamma_2)\cdots(1+\Gamma_{2N-1}\Gamma_{2N})\Gamma_1\cdots\Gamma_{2N}.
\end{eqnarray}
Now, using the JW transformation, we arrive at
\begin{eqnarray}
    {{\cal U}}=(-{\rm i})^N\left[\prod_{j=1}^{N-1}\frac{1+{\rm i}Z_j}{\sqrt{2}}\frac{1+{\rm i}X_jX_{j+1}}{\sqrt{2}}\right]\frac{1+{\rm i}Z_N}{\sqrt{2}}{\sf P}.
\end{eqnarray}
{It is worth noting that the operator $\mathcal{U}$ is unitary and we shall see soon that it acts like the Kramers-Wannier duality operator for a twisted Ising model.} However, it is clear that ${\cal U}$ does not act as some translation operator anymore in the spin basis. To have a better understanding of its action, let us consider the parameter dependent Hamiltonian
\begin{eqnarray}
    {\frak H}(h)=-h\sum_{j=1}^NZ_j-\sum_{j=1}^{N-1}X_jX_{j+1}-{\sf P}X_NX_1,\quad h\in\mathbb{R},
\end{eqnarray}
which reduces to $H$ at $h=1$, i.e. one has ${\frak H}(1)=H$. It is now straightforward to check that, under the action of ${\cal U}$, the Hamiltonian ${\frak H}(h)$ transforms as
\begin{eqnarray}
    {\cal U}{\frak H}(h){\cal U}^{-1}=h~{\frak H}(1/h).
\end{eqnarray}
Therefore, the unitary operator ${\cal U}$ implements the duality between the Hamiltonian ${\frak H}(h)$ and its dual ${\frak H}(1/h)$, which evidently becomes a symmetry at $h=1$. This resembles the familiar Kramers-Wannier (KW) duality of the TFIM. However, the Hamiltonian $H$ for finite $N$ does not exactly describe the TFIM due to the presence of the parity-dependent term ${\sf P}X_NX_1$. To remedy this, we will now consider the TFIM with periodic boundary condition and demonstrate how systematically one can exploit the integrability of $H$ to construct the KW duality operator, among other (non-invertible) symmetries.
 
\section{Non-invertible symmetries and the Kramers-Wannier duality}
\label{subsec:KW-symmetry}
We will now see how the KW duality and other non-invertible symmetries appear in the TFIM on a finite lattice with periodic boundary conditions. To begin with, let us rewrite Hamiltonian in the following way:
\begin{eqnarray}\label{eq:Ising-P}
     H&=&H_{\rm TFIM}+(1-{\sf P})X_1X_N,
\end{eqnarray}
where the $H_{\rm TFIM}$ is the Hamiltonian corresponding to the periodic transverse-field Ising model, as given in \eqref{eq:Ising-H}. As mentioned before, the mutually commuting conserved quantities \eqref{eq:consq} generate the Abelian algebra ${\cal Q}$, every element of which again is a conserved quantity. A particularly important one is the parity ${\sf P}$, which by the very definition, commutes with any $Q\in{\cal Q}$. Notably, ${\sf P}$ also commutes with $H_{\rm TFIM}$. 

However, an arbitrary element from ${\cal Q}$ may not also be a conserved quantity of $H_{\rm TFIM}$. An example is the twisted translation operator ${\cal U}$, which although is an element of ${\cal Q}$, does not commute with $H_{\rm TFIM}$. We shall now identify the subset of ${\cal Q}$, which also commutes with $H_{\rm TFIM}$. To begin with, let us consider an arbitrary element $Q\in{\cal Q}$ and construct
\begin{eqnarray}
    Q_+=Q(1+{\sf P})=(1+{\sf P})Q,
\end{eqnarray}
where we have used the fact that $Q$ and ${\sf P}$ commute among themselves. Notably, $Q_+$ is non-invertible and again is an element of ${\cal Q}$. Now it is easy to check that
\begin{eqnarray}
    \left[Q_+,H_{\rm TFIM}\right]=0,\quad\forall Q\in{\cal Q},
\end{eqnarray}
as $\left[(1+{\sf P})Q, (1-{\sf P})X_1X_N\right]=0$.
Therefore, the TFIM at the criticality comes with a number of non-invertible symmetries, described by the set ${\cal Q}_+=\{Q_+\}\subset{\cal Q}$. 

To illustrate this idea, we will now take up the concrete example of $Q={\cal U}$, and consider the operator
\begin{eqnarray}
    \mathsf{D}:={\cal U}_+={{\cal U}}\left({1+{\sf P}}\right),\quad [\mathsf{D},H_{\rm TFIM}]=0,
\end{eqnarray}
which belongs to the algebra ${\cal Q}_+$. This particular operator is rather special, in the sense that it acts in a local way. However, it is non-invertible, as seen from the construction and therefore cannot act via conjugation. Nevertheless, one can define its action in the following way:
\begin{eqnarray}
    {\sf D}:\alpha\to\beta\quad\text{if}\quad {\sf D}\alpha=\beta{\sf D}.
\end{eqnarray}
Now the action of this operator on the local terms can be expressed as
\begin{eqnarray}
    {\sf D}&:&Z_j\to X_{j}X_{j+1},\quad {\sf D}:X_{j}X_{j+1}\to Z_{j+1}
\end{eqnarray}
with $j=1,\cdots,N$ and $N+1$-th site being identified with the first site itself. It can be easily verified that this is a symmetry of the TFIM with periodic boundary condition and also respects the translation symmetry of the system. The explicit form of the operator $\mathsf{D}$ can be obtained as
\begin{eqnarray}\label{eq:KW-duality-operator}
    {\sf D}=(-{\rm i})^N\left[\prod_{j=1}^{N-1}\frac{1+{\rm i}Z_j}{\sqrt{2}}\frac{1+{\rm i}X_jX_{j+1}}{\sqrt{2}}\right]\frac{1+{\rm i}Z_N}{\sqrt{2}}\left({1+{\sf P}}\right).
\end{eqnarray}
Up to some constant factor, this expression is obtained previously in the context of non-invertible symmetry of the critical Ising model (see Eqn. $237$, of \cite{seiberg2024majorana}, also see Eqn. $3.48$, of \cite{shao2023s}). In particular, this is the non-invertible operator implementing the KW duality, which commutes with the periodic quantum Ising chain at criticality. Therefore we conclude that, the non-invertible KW duality operator of the periodic quantum Ising model, which becomes a symmetry at criticality, can be obtained consistently starting from the quantum inverse-scattering formalism.

\begin{remark}
 One also can construct ${\sf D}_\perp:={\cal U}_-={\cal U}\left({1-{\sf P}}\right)$, which again commutes with the quantum Ising chain at criticality, but now with antiperiodic boundary condition. We emphasize that, our parity-dependent Hamiltonian \eqref{eq:Ising-P} commutes with all the operators ${\cal U},\mathsf{D}:={\cal U}_+,\mathsf{D}_\perp:={\cal U}_-$. Whereas, the periodic and antiperiodic critical Ising Hamiltonians commute only with $\mathsf{D}:={\cal U}_+$ and $\mathsf{D}_\perp:={\cal U}_-$, respectively.     
\end{remark}

\section{Higher conserved charges}
\label{sec:boost}
The commuting transfer matrices give access to an infinite number of conserved charges. We will compute these keeping in mind their source, the Majorana fermion $R$-matrix, and the differences that arise due to this when compared to local, regular $R$-matrices. Extracting these conserved quantities individually by expanding the transfer matrix may not be the best idea all the time. Interestingly, for integrable systems constructed out of regular $R$-matrices, there exists an alternative method that lets one generate the tower of the commuting conserved charges in a more controlled manner. This so-called \textit{boost operator method} involves the \textit{boost operator}, ${\cal B}$, which acts as the generator of the conserved quantities:
\begin{eqnarray}
    \left[{\cal B},\mathbb{Q}_r\right]\cong\mathbb{Q}_{r+1},
\end{eqnarray}
with $\mathbb{Q}_r$ being the conserved charges.
Although, this is a well established method for integrable systems defined over a Hilbert space having tensor product structure \cite{loebbert2016lectures}, a suitable generalization of this for our case is not very obvious. Therefore, we will briefly go over the required derivation.

To begin with, we will consider an infinitely extended chain of Majorana modes. The local Hamiltonian density $h_{j,j+1}$ reads as
\begin{eqnarray}
    h_{j,j+1}={\rm i}\Gamma_j\Gamma_{j+1}=P^-_{j,j+1}\dot{R}_{j,j+1}(0),\quad \dot{R}_{ij}(0)=\frac{{\rm d}R_{ij}(\lambda)}{{\rm d}\lambda}\Bigg|_{\lambda=0}.
\end{eqnarray}
Although the $R$-matrix is non-local, the Hamiltonian density $h_{j,j+1}$ is local in both the fermion and the qubit pictures. Next, we consider the Yang-Baxter equation
\begin{eqnarray}
    R_{a,j}(\lambda)R_{a,j+1}(\mu)R_{j,j+1}(\mu-\lambda)=R_{j,j+1}(\mu-\lambda)R_{a,j+1}(\mu)R_{a,j}(\lambda),
\end{eqnarray}
where $j$ and $a$ are the physical and auxiliary indices, respectively. Differentiating with respect to $\mu$ and setting $\mu=\lambda$ results in the following:
\begin{eqnarray}
    &&R_{a,j}(\lambda)R_{a,j+1}(\lambda)P^-_{j,j+1}h_{j,j+1}+R_{a,j}(\lambda)\dot{R}_{a,j+1}(\lambda)P^-_{j,j+1}=\nonumber\\
    &&\quad\quad\quad\quad\quad P^-_{j,j+1}h_{j,j+1}R_{a,j+1}(\lambda)R_{a,j}(\lambda)+P^-_{j,j+1}\dot{R}_{a,j+1}(\lambda)R_{a,j}(\lambda).
\end{eqnarray}
Ideally, now one should use the properties of the permutation operator to shuffle the indices. However, the operator $P^-_{j,j+1}$ not quite the exact permutation operator, as seen earlier. A simple observation reveals that we rather have
\begin{eqnarray}
    R_{a,j+1}(\lambda)P^-_{j,j+1}=-P^-_{j,j+1}R_{a,j}(\lambda),
\end{eqnarray}
where there is an extra negative sign on the right hand side. Nevertheless, since we have to move the permutation operator through two such $R$-matrices, the negative signs cancel and we arrive at
\begin{eqnarray}
    R_{a,j}(\lambda)R_{a,j+1}(\lambda)P^-_{j,j+1}=P^-_{j,j+1}R_{a,j+1}(\lambda)R_{a,j}(\lambda).
\end{eqnarray}
Therefore, we arrive at the following important relation:
\begin{eqnarray}
    \left[h_{j,j+1},R_{a,j+1}(\lambda)R_{a,j}(\lambda)\right]=R_{a,j+1}(\lambda)\dot{R}_{a,j}(\lambda)-\dot{R}_{a,j+1}(\lambda){R}_{a,j}(\lambda).
\end{eqnarray}
Now, multiplying the above equation from left by $\prod_{k=\infty}^{j+2}R_{a,k}(\lambda)$ and from right by $\prod_{k=j-1}^{-\infty}R_{a,k}(\lambda)$, we obtain
\begin{eqnarray}
    &&\left[h_{j,j+1},T_a(\mu)\right]=\nonumber\\ &&\prod_{k=\infty}^{j+1}R_{a,k}(\lambda)\left[\dot{R}_{a,j}(\lambda)\right]\prod_{k=j-1}^{-\infty} R_{a,k}(\lambda)-\prod_{k=\infty}^{j+2}R_{a,k}(\lambda)\left[\dot{R}_{a,j+1}(\lambda)\right]\prod_{k=j}^{-\infty} R_{a,k}(\lambda).
\end{eqnarray}
Notably, we made use of the fact that $h_{j,j+1}$ commutes with any $R_{a,k}(\lambda)$ with $k\neq j,j+1$. If we now multiply both the sides by $j$ and sum over $j$ from $-\infty$ to $\infty$, followed by tracing over the auxiliary index $a$, we arrive at
\begin{eqnarray}
    \frac{{\rm d}\tau(\lambda)}{{\rm d}\lambda}=[{\cal B},\tau(\lambda)],
\end{eqnarray}
where the boost operator ${\cal B}$ is defined to be ${\cal B}=\sum_{j}j~h_{j,j+1}$. This lets us generate the tower of conserved quantities \cite{loebbert2016lectures}, starting from the Hamiltonian ${\mathbb Q}_2=H$. One immediately finds out that the next conserved charge to be
\begin{eqnarray}
    \mathbb{Q}_3={\rm i}\sum_j\Gamma_j\Gamma_{j+2}.
\end{eqnarray}
By mathematical induction, we can now prove that an arbitrary conserved charge is given by\footnote{{These charges were derived from a generalization of the Jordan-Wigner transformation \cite{Minami_2017,Minami_2021,Minami_2025}. These methods also give the conserved quantities corresponding to other free-fermionic spin chains making it a universal method.}}
\begin{eqnarray}\label{eq:general-Qp}
    \mathbb{Q}_p={\rm i}\sum_j \Gamma_j\Gamma_{j+p-1}.
\end{eqnarray}
Suppose, it is given that, we have the conserved quantities ${\mathbb Q}_{\{r/r+1\}}={\rm i}\sum_k\Gamma_k\Gamma_{\{k+r-1/k+r\}}$. The next conserved quantity $\mathbb{Q}_{r+2}={\rm i}\sum_j\Gamma_j\Gamma_{j+r+1}$ should be obtained from $[\mathcal{B},\mathbb{Q}_{r+1}]$. We expand the above commutator as follows:
\begin{eqnarray}   
[\mathcal{B},\mathbb{Q}_{r+1}]&=&-\sum_{k,l}k[\Gamma_k\Gamma_{k+1},\Gamma_l\Gamma_{l+r}]\nonumber\\
    &=&-\sum_k k([\Gamma_k\Gamma_{k+1},\Gamma_k\Gamma_{k+r}]+[\Gamma_k\Gamma_{k+1},\Gamma_{k-r}\Gamma_k]\nonumber\\ &&\quad\quad\quad+[\Gamma_k\Gamma_{k+1},\Gamma_{k+1}\Gamma_{k+r+1}]+[\Gamma_k\Gamma_{k+1},\Gamma_{k-r+1}\Gamma_{k+1}]).
\end{eqnarray}
Using the properties of the Majorana fermions, we obtain
\begin{eqnarray}
    [{\cal B},\mathbb{Q}_{r+1}]&=&-2\sum_k k\left(-\Gamma_{k+1}\Gamma_{k+r}+\Gamma_{k+1}\Gamma_{k-r}+\Gamma_k\Gamma_{k+r+1}-\Gamma_k\Gamma_{k-r+1} \right)\nonumber\\
    &=&2{\rm i}r\mathbb{Q}_{r}+2r\sum_{k}\Gamma_k\Gamma_{k+r+1}.
\end{eqnarray}
Thus, we can extract the conserved charge $\mathbb{Q}_{r+2}={\rm i}\sum_j\Gamma_j\Gamma_{j+r+1}$ from the commutator $[{\cal B},{\mathbb Q}_{r+1}]$, as promised. Notably, all the charges are quadratic in $\Gamma$'s. Since the complex fermions, $c$'s, are linear in $\Gamma$'s, when expressed in terms of them, the conserved charges, $\mathbb{Q}_r$'s, remain quadratic. In sharp contrast to this, the $\mathbb{Q}_r$'s are not guaranteed to be quadratic in the spin basis, owing to the non-local nature of the JW transformation \eqref{eq:JW-transform}. 

It should also be noted that such infinite sets of conserved charges for the quantum Ising model were found by other indirect methods in \cite{Grady1982}. These approaches use the commuting property of the transfer matrices of the ice-type eight vertex models. In particular they arrive at the answers by using a mapping between the $XY$ model and the Ising model. While the results in \cite{Grady1982} are written in terms of spin operators, the lattice versions in terms of Majorana fermions, using the methods in \cite{Grady1982}, can be found in Eq. 3.6 of \cite{Fagotti_2013}. The continuum versions of these operators are in Eq. 6 of \cite{essler2015generalized}. We briefly discuss these results now. Following \cite{essler2015generalized}, where the authors have the lattice conserved quantities (with $h=1$ for the criticality)
\begin{eqnarray}
    {\cal I}_n^+&=&\frac{{\rm i}J}{2}\sum_{j}\sum_{\sigma=\pm 1}\Gamma_{2j}\left[\Gamma_{2j+2n\sigma+1}-\Gamma_{2j+2n\sigma-1}\right],\nonumber\\
    {\cal I}_{n-1}^-&=&-\frac{{\rm i}J}{2}\sum_j\left[\Gamma_{2j}\Gamma_{2j+2n}+\Gamma_{2j-1}\Gamma_{2j+2n-1}\right],
\end{eqnarray}
we can identify our findings in Eq. \eqref{eq:general-Qp} as
\begin{eqnarray}
    {\cal I}_n^+&=&\frac{J}{2}\left({\mathbb Q}_{2n+2}-{\mathbb Q}_{2n}\right),\nonumber\\
    {\cal I}_{n-1}^-&=&-\frac{J}{2}{\mathbb Q}_{2n+1}.
\end{eqnarray}

\begin{remark}
    It is well known \cite{pfeuty1970one} that writing the TFIM Hamiltonian in terms of the complex fermions yields a quadratic Hamiltonian of the form
    \begin{eqnarray}
H=\sum_{i,j}\left(c_i^{\dagger}A_{ij}c_j+c_i^{\dagger}B_{ij}c_j^{\dagger}+{\rm h.c}\right).
    \end{eqnarray}
    Such a Hamiltonian can always be diagonalized in a suitable basis and subsequently can be expressed as (up to a constant)
    \begin{eqnarray}
        H=\sum_k\Lambda_k\eta_k^{\dagger}\eta_k.
    \end{eqnarray}
    The quasi-particle occupation number operators, $\hat{n}_k:=\eta_k^{\dagger}\eta_k$'s commute with the Hamiltonian and thus represent conserved quantities. However, in most of the cases, such operators are highly non-local in the real space. On the other hand, by the boost operator technique, we directly obtain the local and quasi-local conserved quantities. Nevertheless, we expect the conserved quantities ${\mathbb Q}_r$'s to have well-defined representations in terms of the $\hat{n}_k$'s.
\end{remark}

\section{Conclusion}
\label{sec:conclusion}
Integrability in quantum theory has long been debated \cite{Caux_2011}. Some of the early definitions included all systems that are exactly solvable \cite{Baxter:1982zz}. A more concrete notion was achieved through the quantum inverse scattering method \cite{Korepin1993QuantumIS} which produces an infinite number of conserved quantities for the Hamiltonian. The starting point for this approach is the Yang-Baxter equation. We call the systems obtained through this method as Yang-Baxter integrable. In this work we have developed the $R$-matrix [Yang-Baxter solution] that realizes the above program for the critical transverse field Ising model in one dimension. This $R$-matrix is quite different from the traditional $R$-matrices for say the six-vertex and eight-vertex models. This could also be a reason why this $R$-matrix has eluded discovery in all these years. So in the process we have also broadened the scope of applicability of the Yang-Baxter equation. Following all this the associated conserved quantities [higher conserved charges] are also computed systematically using the boost operator method. Interestingly, we find the Kramers-Wannier duality transformation, which turns into a non-invertible symmetry at criticality, among the conserved charges. Thus the $R$-matrix for the critical transverse field Ising model can be seen as another source of this duality. 

This work lays the foundation for several follow ups. Some of them are listed here:
\begin{enumerate}
    \item The algebraic Bethe ansatz method needs to be developed for the types of $R$-matrices introduced in this work. This will help us construct the spectrum and correlations functions of this model.
    \item The existence of this $R$-matrix implies that we can construct an integrable quantum circuit \cite{Vanicat2017IntegrableTL} to simulate the transverse field Ising model on a quantum computer. This circuit is expected to slow down thermalization due to the many conserved quantities.
    \item Majorana fermions have also been used to construct solutions to the tetrahedron equations and other higher simplex equations \cite{PADMANABHAN2025116865,singh2024unitarytetrahedronquantumgates}. It would interesting to construct the local two dimensional quantum models obtained from these solutions and see their relation to the three dimensional Ising model.
    \item At criticality the transverse field Ising model coincides with the critical Kitaev chain modeling a $p$-wave superconductor \cite{Kitaev_2001}. We know that this system hosts Majorana zero modes that are not localized at the edges. Our initial numerical computations suggest that there are many zero modes at this point. We hope to account for them using the conserved quantities obtained here and see which among them are stable and capable of quantum information processing.
    \item {The QISM gives a new perspective on KW dualities and, in general, non-invertible symmetries. These new symmetries have also been obtained using fusion categories \cite{Zhang_2025}. These two approaches have to be related and so it would be interesting to obtain this connection.}
\end{enumerate}

\section*{Acknowledgments}
PP, AS and TJ thank Kush Saha and Somnath Maity for useful discussions.

VK is funded by the U.S. Department of Energy, Office of Science, National Quantum Information Science Research Centers, Co-Design Center for Quantum Advantage ($C^2QA$) under Contract No. DE-SC0012704.

\appendix

\section{The operator $P^+$}
\label{app:P+}
It is possible to define an alternate permutation-like operator $P^+$ in place of $P^-$ in \eqref{eq:P-Majorana} as
\begin{eqnarray}\label{eq:P-Majorana-+}
    P^+_{jk} = \frac{\gamma_j+\gamma_k}{\sqrt{2}}.
\end{eqnarray}
Here too, the inverse is itself
\begin{eqnarray}\label{eq:P-inverse-Majorana-+}
   (P^+_{jk})^{-1}=P^+_{jk} = \frac{\gamma_j+\gamma_k}{\sqrt{2}}. 
\end{eqnarray}
However in this case it satisfies the braided version of a modified YBE
$$ P^+_{ij}P^+_{jk}P^+_{ij}=-P^+_{jk}P^+_{ij}P^+_{jk} = P^-_{ik},  $$
known as the {\it anti-YBE} \cite{PADMANABHAN2024116664}. It does not satisfy the non-braided YBE or non-braided anti-YBE. Due to this fact we cannot use this operator for converting a spectral parameter dependent braided YBO into a non-braided YBO. Thus its use for obtaining integrable models this way is limited.

Just as the $P^-$ operator it satisfies a far-anticommutativity relation
\begin{eqnarray}
    P^+_{ij}P^+_{kl} = -P^+_{kl}P^+_{ij}, 
\end{eqnarray}
instead of the far-commutativity relation. Both these relations distinguishes it from the usual permutation operator. The consequence is also seen in the way this permutation interchanges other Majorana fermion operators. The action is by conjugation
\begin{eqnarray}\label{eq:ptilde+-action-majoranas}
    &P^+_{jk}\gamma_jP^+_{jk} = \gamma_k~;~ P^+_{jk}\gamma_k P^+_{jk} = \gamma_j, & \nonumber \\
    &  P^+_{jk}\gamma_l P^+_{jk} = -\gamma_l,~~~\textrm{when}~l\neq j,k.   &
\end{eqnarray}



{
\section{Odd Majorana modes}\label{app:oddMaj}
We start with the $RTT$-relation, which also can be written as
\begin{eqnarray}
    R_{a,b}(\lambda-\mu)T_a(\lambda)T_b(\mu)R_{a,b}(\lambda-\mu)=(1+f_{\lambda-\mu}^2)T_b(\mu)T_a(\lambda).
\end{eqnarray}
If we expand the expression on the left side of the above equation
\begin{eqnarray}
    &R_{a,b}(\lambda-\mu)T_a(\lambda)T_b(\mu)R_{a,b}(\lambda-\mu)&\nonumber\\
    =&(\gamma_a+f_{\lambda-\mu}\gamma_b)\left[\tau(\lambda)+\gamma_a\sigma(\lambda)\right]\left[\tau(\mu)+\gamma_b\sigma(\mu)\right](\gamma_a+f_{\lambda-\mu}\gamma_b),&
\end{eqnarray}
it becomes a linear combination of the eight possible terms, namely
\begin{eqnarray}
&\tau(\lambda)\tau(\mu),~\gamma_a\gamma_b\tau(\lambda)\tau(\mu),~\sigma(\lambda)\sigma(\mu),~\gamma_a\gamma_b\sigma(\lambda)\sigma(\mu),&\nonumber\\
&\gamma_a\tau(\lambda)\sigma(\mu),~\gamma_b\tau(\lambda)\sigma(\mu),~\gamma_a\sigma(\lambda)\tau(\mu),~\gamma_b\sigma(\lambda)\tau(\mu).&
\end{eqnarray}
This is true algebraically, regardless the representation of the Majorana modes. Since taking trace is a linear operation, it is advisable to check which of the above eight terms give non-zero results after performing partial trace. However, this demands choosing a representation, which is different for even and odd number of Majorana modes. For even Majorana modes, the representation is provided by the Jordan-Wigner transformation \eqref{eq:JW-transform}. As a consequence, $\tau(x)$, which contains even number of $\gamma_j,~j\neq a,b$, acts trivially on the auxiliary indices $a,b$. However, $\sigma(x)$ carries odd number of $\gamma_j,~j\neq a,b$ and acts non-trivially on the auxiliary indices. We can then write
\begin{eqnarray}
    \sigma(x)=Z_a\sigma'(x),
\end{eqnarray}
where $\sigma'(x)$ now acts trivially on the auxiliary indices. One can perform the partial trace over the auxiliary indices $a,b$ and see that, the only non-zero results are
\begin{eqnarray}
    {\rm tr}_{a,b}[\tau(\lambda)\tau(\mu)]=\tau(\lambda)\tau(\mu),\quad {\rm tr}_{a,b}[\sigma(\lambda)\sigma(\mu)]=\sigma'(\lambda)\sigma'(\mu).
\end{eqnarray}
This leads to the commutativity of the transfer matrices
\begin{eqnarray}
    [\tau(\lambda),\tau(\mu)]=0,\quad \tau(\lambda)={\rm tr}_a~T_a(\lambda)={\rm tr}_b~T_b(\lambda).
\end{eqnarray}

To treat an odd number of Majoranas, we shall work with the following Jordan-Wigner transformation:
\begin{eqnarray}    &\gamma_a=X_a,\quad\gamma_b=Z_a\left[\prod\limits_{k=1}^{N+1}Z_k\right]X_b&\nonumber\\
    &\gamma_{2j-1}=Z_a\left[\prod\limits_{k=1}^{j-1}Z_k\right]X_j,\quad \gamma_{2j}=Z_a\left[\prod\limits_{k=1}^{j-1}Z_k\right]Y_j,\quad \gamma_{2N+1}=Z_a\left[\prod\limits_{k=1}^{N}Z_k\right]X_{N+1}.&
\end{eqnarray}
In sharp contrast to the case of even Majorana modes, $\tau(x)$ now contains odd number of $\gamma_j,~j\neq a,b$ and acts non-trivially on the auxiliary indices. Whereas, $\sigma(x)$ has even number of $\gamma_j,~j\neq a,b$, thus acting trivially on $a,b$. We now can write
\begin{eqnarray}
    \tau(x)=Z_a\tau'(x),
\end{eqnarray}
where $\tau'(x)$ acts trivially on $a,b$. This is the reason why a naive trace over the auxiliary index $a$ gives zero for odd ${\cal N}$ in Eqn.\eqref{eq:even-odd TM}. An analysis similar to the case of even Majoranas leads to
\begin{eqnarray}
    [\tau'(\lambda),\tau'(\mu)]=0.
\end{eqnarray}
Therefore, the operator $\tau'(x)$ can be considered as the appropriate transfer matrix in this scenario. However, a straightforward attempt to extract $\tau'(\lambda)$ from the monodromy matrices by taking the partial trace fails
\begin{eqnarray}
    {\rm tr}_a[T_a(\lambda)]=0,\quad {\rm tr}_b[T_b(\lambda)]=Z_a\tau'(\lambda),
\end{eqnarray}
as we saw earlier. Nonetheless, we can obtain $\tau'(\lambda)$ by the trick
\begin{eqnarray}
    \tau'(\lambda)={\rm tr}_a[Z_aT_a(\lambda)]={\rm tr}_b[Z_aT_b(\lambda)].
\end{eqnarray}
This motivates the introduction of the \textit{supertrace} as
\begin{eqnarray}
    {\rm str}_{a(b);~{\cal N}}[~\cdot~]={\rm tr}_{a(b)}[Z_a^{\cal N}~\cdot~],
\end{eqnarray}
for ${\cal N}$ both even and odd. For even ${\cal N}$, the supertrace reduces to the ordinary trace, as expected. The commuting transfer matrices now can be extracted uniquely as
\begin{eqnarray}
    {\rm str}_{a;~{\cal N}}[T_a(\lambda)]={\rm str}_{b;~{\cal N}}[T_b(\lambda)].
\end{eqnarray}
Since the monodromy operator is bosonic and fermionic in nature for even and odd number of Majorana modes, respectively, the idea of introducing a notion of supertrace is not very surprising. Naively speaking, the operator $Z_a^{\cal N}$ behaves like a higher-dimensional analogue of $(-1)^F$, where $F$ denotes the fermionic parity and is $0$ for bosonic and $1$ for fermionic variables.
}

\bibliographystyle{acm}
\normalem
\bibliography{refs}

\begin{thebibliography}{10}

\bibitem{aasen2020topological}
{\sc Aasen, D., Fendley, P., and Mong, R.~S.}
\newblock Topological defects on the lattice: dualities and degeneracies.
\newblock {\em arXiv preprint arXiv:2008.08598\/} (2020).

\bibitem{aasen2016topological}
{\sc Aasen, D., Mong, R.~S., and Fendley, P.}
\newblock Topological defects on the lattice: I. the ising model.
\newblock {\em Journal of Physics A: Mathematical and Theoretical 49}, 35 (2016), 354001.

\bibitem{AUYANG1987219}
{\sc Au-Yang, H., McCoy, B.~M., Perk, J.~H., Tang, S., and Yan, M.-L.}
\newblock Commuting transfer matrices in the chiral potts models: Solutions of star-triangle equations with genus>1.
\newblock {\em Physics Letters A 123}, 5 (1987), 219--223.

\bibitem{au1989onsager}
{\sc Au-Yang, H., and Perk, J.~H.}
\newblock Onsager's star-triangle equation: Master key to integrability.
\newblock In {\em Integrable systems in quantum field theory and statistical mechanics}. Elsevier, 1989, pp.~57--94.

\bibitem{AuYang1996TheMF}
{\sc Au-Yang, H., and Perk, J. H.~H.}
\newblock The many faces of the chiral potts model.
\newblock {\em International Journal of Modern Physics B 11\/} (1996), 11--26.

\bibitem{AuYang2016About3Y}
{\sc Au-Yang, H., and Perk, J. H.~H.}
\newblock About 30 years of integrable chiral potts model, quantum groups at roots of unity and cyclic hypergeometric functions.
\newblock {\em arXiv: Mathematical Physics\/} (2016).

\bibitem{Barouch1971}
{\sc Barouch, E., and McCoy, B.~M.}
\newblock Statistical mechanics of the $xy$ model. ii. spin-correlation functions.
\newblock {\em Phys. Rev. A 3\/} (Feb 1971), 786--804.

\bibitem{Batchelor_2015}
{\sc Batchelor, M.~T., and Zhou, H.-Q.}
\newblock Integrability versus exact solvability in the quantum rabi and dicke models.
\newblock {\em Physical Review A 91}, 5 (May 2015).

\bibitem{BAXTER1988138}
{\sc Baxter, R., Perk, J., and Au-Yang, H.}
\newblock New solutions of the star-triangle relations for the chiral potts model.
\newblock {\em Physics Letters A 128}, 3 (1988), 138--142.

\bibitem{BAXTER1972193}
{\sc Baxter, R.~J.}
\newblock Partition function of the eight-vertex lattice model.
\newblock {\em Annals of Physics 70}, 1 (1972), 193--228.

\bibitem{Baxter:1982zz}
{\sc Baxter, R.~J.}
\newblock {\em {Exactly solved models in statistical mechanics}}.
\newblock 1982.

\bibitem{BERNARD_1993}
{\sc Bernard, D.}
\newblock An introduction to yangian symmetries.
\newblock {\em International Journal of Modern Physics B 07}, 20n21 (Sept. 1993), 3517–3530.

\bibitem{Braak2011}
{\sc Braak, D.}
\newblock Integrability of the rabi model.
\newblock {\em Phys. Rev. Lett. 107\/} (Aug 2011), 100401.

\bibitem{Bravyi_2002}
{\sc Bravyi, S.~B., and Kitaev, A.~Y.}
\newblock Fermionic quantum computation.
\newblock {\em Annals of Physics 298}, 1 (May 2002), 210–226.

\bibitem{Caux_2011}
{\sc Caux, J.-S., and Mossel, J.}
\newblock Remarks on the notion of quantum integrability.
\newblock {\em Journal of Statistical Mechanics: Theory and Experiment 2011}, 02 (Feb. 2011), P02023.

\bibitem{Delfino2003IntegrableFT}
{\sc Delfino, G.}
\newblock Integrable field theory and critical phenomena: The ising model in a magnetic field.
\newblock {\em Journal of Physics A 37\/} (2003).

\bibitem{Delfino_1995}
{\sc Delfino, G., and Mussardo, G.}
\newblock The spin-spin correlation function in the two-dimensional ising model in a magnetic field at t = tc.
\newblock {\em Nuclear Physics B 455}, 3 (Sept. 1995), 724–758.

\bibitem{Drinfeld1988}
{\sc Drinfel'd, V.~G.}
\newblock Quantum groups.
\newblock {\em Journal of Soviet Mathematics 41}, 2 (1988), 898--915.

\bibitem{Elliott1970}
{\sc Elliott, R.~J., Pfeuty, P., and Wood, C.}
\newblock Ising model with a transverse field.
\newblock {\em Phys. Rev. Lett. 25\/} (Aug 1970), 443--446.

\bibitem{esslerfrahmgohmannklumperkorepin2005}
{\sc Essler, F. H.~L., Frahm, H., Göhmann, F., Klümper, A., and Korepin, V.~E.}
\newblock {\em The One-Dimensional Hubbard Model}.
\newblock Cambridge University Press, 2005.

\bibitem{essler2015generalized}
{\sc Essler, F. H.~L., Mussardo, G., and Panfil, M.}
\newblock Generalized gibbs ensembles for quantum field theories.
\newblock {\em Physical Review A 91}, 5 (2015), 051602.

\bibitem{faddeev1996algebraic}
{\sc Faddeev, L.}
\newblock How algebraic bethe ansatz works for integrable model.
\newblock {\em arXiv preprint hep-th/9605187\/} (1996).

\bibitem{Faddeev1984SpectrumAS}
{\sc Faddeev, L.~D., and Takhtadzhyan, L.~A.}
\newblock Spectrum and scattering of excitations in the one-dimensional isotropic heisenberg model.
\newblock {\em Journal of Soviet Mathematics 24\/} (1984), 241--267.

\bibitem{Fagotti_2013}
{\sc Fagotti, M., and Essler, F. H.~L.}
\newblock Reduced density matrix after a quantum quench.
\newblock {\em Physical Review B 87}, 24 (June 2013).

\bibitem{Fonseca2001IsingFT}
{\sc Fonseca, P., and Zamolodchikov, A.~B.}
\newblock Ising field theory in a magnetic field: Analytic properties of the free energy.
\newblock {\em Journal of Statistical Physics 110\/} (2001), 527--590.

\bibitem{Susskind1978}
{\sc Fradkin, E., and Susskind, L.}
\newblock Order and disorder in gauge systems and magnets.
\newblock {\em Phys. Rev. D 17\/} (May 1978), 2637--2658.

\bibitem{frohlich2004kramers}
{\sc Fr{\"o}hlich, J., Fuchs, J., Runkel, I., and Schweigert, C.}
\newblock Kramers-wannier duality from conformal defects.
\newblock {\em Physical review letters 93}, 7 (2004), 070601.

\bibitem{Grady1982}
{\sc Grady, M.}
\newblock Infinite set of conserved charges in the ising model.
\newblock {\em Phys. Rev. D 25\/} (Feb 1982), 1103--1113.

\bibitem{Hecht1967}
{\sc Hecht, R.}
\newblock Correlation functions for the two-dimensional ising model.
\newblock {\em Phys. Rev. 158\/} (Jun 1967), 557--561.

\bibitem{Henkel1989TheTI}
{\sc Henkel, M., and Saleur, H.}
\newblock The two-dimensional ising model in the magnetic field: a numerical check of zamolodchikov's conjecture.
\newblock {\em Journal of Physics A 22\/} (1989).

\bibitem{Houtappel1950}
{\sc {Houtappel}, R.~M.~F.}
\newblock {Order-disorder in hexagonal lattices}.
\newblock {\em Physica 16}, 5 (May 1950), 425--455.

\bibitem{ITS1990752}
{\sc Its, A., Izergin, A., Korepin, V., and Novokshenov, V.}
\newblock Temperature autocorrelations of the transverse ising chain at the critical magnetic field.
\newblock {\em Nuclear Physics B 340}, 2 (1990), 752--758.

\bibitem{Its1993}
{\sc Its, A.~R., Izergin, A.~G., Korepin, V.~E., and Slavnov, N.~A.}
\newblock Temperature correlations of quantum spins.
\newblock {\em Phys. Rev. Lett. 70\/} (Mar 1993), 1704--1706.

\bibitem{Ivanov2001}
{\sc Ivanov, D.~A.}
\newblock Non-abelian statistics of half-quantum vortices in $\mathit{p}$-wave superconductors.
\newblock {\em Phys. Rev. Lett. 86\/} (Jan 2001), 268--271.

\bibitem{Izergin1984TheQI}
{\sc Izergin, A.~G., and Korepin, V.~E.}
\newblock The quantum inverse scattering method approach to correlation functions.
\newblock {\em Communications in Mathematical Physics 94\/} (1984), 67--92.

\bibitem{Jones1990}
{\sc Jones, V. F.~R.}
\newblock {\em Baxterization}.
\newblock Springer US, Boston, MA, 1990, pp.~5--11.

\bibitem{Kauffman_2018}
{\sc Kauffman, L.~H.}
\newblock Majorana fermions and representations of the braid group.
\newblock {\em International Journal of Modern Physics A 33}, 23 (Aug. 2018), 1830023.

\bibitem{Kauffman_2016}
{\sc Kauffman, L.~H., and Lomonaco, S.~J.}
\newblock Braiding with majorana fermions.
\newblock In {\em Quantum Information and Computation IX\/} (May 2016), E.~Donkor and M.~Hayduk, Eds., SPIE.

\bibitem{Kaufman1949}
{\sc Kaufman, B.}
\newblock Crystal statistics. ii. partition function evaluated by spinor analysis.
\newblock {\em Phys. Rev. 76\/} (Oct 1949), 1232--1243.

\bibitem{sedrakyan-1}
{\sc Khachatryan, S., and Sedrakyan, A.}
\newblock Characteristics of two-dimensional lattice models from a fermionic realization: Ising and $xyz$ models.
\newblock {\em Phys. Rev. B 80\/} (Sep 2009), 125128.

\bibitem{Kirillov1986ExactSO}
{\sc Kirillov, A.~N., and Reshetikhin, N.}
\newblock Exact solution of the heisenberg xxz model of spin s.
\newblock {\em Journal of Soviet Mathematics 35\/} (1986), 2627--2643.

\bibitem{Kirillov1987ExactSO}
{\sc Kirillov, A.~N., and Reshetikhin, N.}
\newblock Exact solution of the integrable xxz heisenberg model with arbitrary spin. i. the ground state and the excitation spectrum.
\newblock {\em Journal of Physics A 20\/} (1987), 1565--1585.

\bibitem{Kitaev_2001}
{\sc Kitaev, A.~Y.}
\newblock Unpaired majorana fermions in quantum wires.
\newblock {\em Physics-Uspekhi 44}, 10S (oct 2001), 131.

\bibitem{Koo_1994}
{\sc Koo, W., and Saleur, H.}
\newblock Representations of the virasoro algebra from lattice models.
\newblock {\em Nuclear Physics B 426}, 3 (Sept. 1994), 459–504.

\bibitem{Korepin1984CorrelationFO}
{\sc Korepin, V.~E.}
\newblock Correlation functions of the one-dimensional bose gas in the repulsive case.
\newblock {\em Journal of Soviet Mathematics 31\/} (1984), 3344--3351.

\bibitem{Korepin1993QuantumIS}
{\sc Korepin, V.~E., Bogoliubov, N.~M., and Izergin, A.~G.}
\newblock Quantum inverse scattering method and correlation functions.
\newblock Cambridge university press.

\bibitem{Korepin1994CorrelationFO}
{\sc Korepin, V.~E., Izergin, A.~G., Essler, F. H.~L., and Uglov, D.~B.}
\newblock Correlation function of the spin 1/2 xxx antiferromagnet.
\newblock {\em Physics Letters B\/} (1994), 22--24.

\bibitem{KW1941-1}
{\sc Kramers, H.~A., and Wannier, G.~H.}
\newblock Statistics of the two-dimensional ferromagnet. part i.
\newblock {\em Phys. Rev. 60\/} (Aug 1941), 252--262.

\bibitem{KW1941-2}
{\sc Kramers, H.~A., and Wannier, G.~H.}
\newblock Statistics of the two-dimensional ferromagnet. part ii.
\newblock {\em Phys. Rev. 60\/} (Aug 1941), 263--276.

\bibitem{Kulish_2008}
{\sc Kulish, P.~P., Manojlovic, N., and Nagy, Z.}
\newblock Quantum symmetry algebras of spin systems related to temperley–lieb r-matrices.
\newblock {\em Journal of Mathematical Physics 49}, 2 (Feb. 2008).

\bibitem{loebbert2016lectures}
{\sc Loebbert, F.}
\newblock Lectures on yangian symmetry.
\newblock {\em Journal of Physics A: Mathematical and Theoretical 49}, 32 (2016), 323002.

\bibitem{Maillet2006CorrelationFO}
{\sc Maillet, J.~M.}
\newblock Correlation functions of the xxz heisenberg spin chain: Bethe ansatz approach.

\bibitem{Majid_1995}
{\sc Majid, S.}
\newblock {\em Foundations of Quantum Group Theory}.
\newblock Cambridge University Press, 1995.

\bibitem{Potts_PPMartin}
{\sc Martin, P.}
\newblock {\em Potts Models and Related Problems in Statistical Mechanics}.
\newblock WORLD SCIENTIFIC, 1991.

\bibitem{MARCIO2024116610}
{\sc Martins, M.}
\newblock On the equivalence between n-state spin and vertex models on the square lattice.
\newblock {\em Nuclear Physics B 1005\/} (2024), 116610.

\bibitem{marcio-1}
{\sc {Martins}, M.~J.}
\newblock {Embedding integrable spin models in solvable vertex models on the square lattice}.
\newblock {\em Nuclear Physics B 1013\/} (Apr. 2025), 116849.

\bibitem{MCCOY198335}
{\sc McCoy, B.~M., Perk, J.~H., and Shrock, R.~E.}
\newblock Time-dependent correlation functions of the transverse ising chain at the critical magnetic field.
\newblock {\em Nuclear Physics B 220}, 1 (1983), 35--47.

\bibitem{MCCOY19879}
{\sc McCoy, B.~M., Perk, J.~H., Tang, S., and {Chih-Han Sah}}.
\newblock Commuting transfer matrices for the four-state self-dual chiral potts model with a genus-three uniformizing fermat curve.
\newblock {\em Physics Letters A 125}, 1 (1987), 9--14.

\bibitem{McCoy:1977er}
{\sc McCoy, B.~M., Tracy, C.~A., and Wu, T.~T.}
\newblock {Two-Dimensional Ising Model as an Exactly Solvable Relativistic Quantum Field Theory: Explicit Formulas for n Point Functions}.
\newblock {\em Phys. Rev. Lett. 38\/} (1977), 793--796.

\bibitem{McCoyWu+1973}
{\sc McCoy, B.~M., and Wu, T.~T.}
\newblock {\em The Two-Dimensional Ising Model}.
\newblock Harvard University Press, Cambridge, MA and London, England, 1973.

\bibitem{mills_ascher_jaffee_1971}
{\sc Mills, R., Ascher, E., and Jaffee, R.}
\newblock {\em Critical phenomena in alloys, magnets, and superconductors. Battelle Institute materials science colloquia, Geneva and Gstaad, Switzerland, September 7--12, 1970}.
\newblock McGraw-Hill Book Company., 1971.

\bibitem{Minami_2017}
{\sc Minami, K.}
\newblock Infinite number of solvable generalizations of xy-chain, with cluster state, and with central charge c = m /2.
\newblock {\em Nuclear Physics B 925\/} (Dec. 2017), 144–160.

\bibitem{Minami_2021}
{\sc Minami, K.}
\newblock Onsager algebra and algebraic generalization of jordan-wigner transformation.
\newblock {\em Nuclear Physics B 973\/} (Dec. 2021), 115599.

\bibitem{Minami_2025}
{\sc Minami, K.}
\newblock Conserved charges of series of solvable lattice models.
\newblock {\em Nuclear Physics B 1012\/} (Mar. 2025), 116844.

\bibitem{Mussardo-book}
{\sc Mussardo, G.}
\newblock {\em Statistical Field Theory: An Introduction to Exactly Solved Models in Statistical Physics}.
\newblock Oxford University Press, 03 2020.

\bibitem{Onsager1944}
{\sc Onsager, L.}
\newblock Crystal statistics. i. a two-dimensional model with an order-disorder transition.
\newblock {\em Phys. Rev. 65\/} (Feb 1944), 117--149.

\bibitem{PADMANABHAN2024116664}
{\sc Padmanabhan, P., and Korepin, V.~E.}
\newblock Solving the yang-baxter, tetrahedron and higher simplex equations using clifford algebras.
\newblock {\em Nuclear Physics B\/} (2024), 116664.

\bibitem{PADMANABHAN2025116865}
{\sc Padmanabhan, P., and Korepin, V.~E.}
\newblock Majorana fermions solve the tetrahedron equations as well as higher simplex equations.
\newblock {\em Nuclear Physics B\/} (2025), 116865.

\bibitem{Peierls_1936}
{\sc Peierls, R.}
\newblock On ising’s model of ferromagnetism.
\newblock {\em Mathematical Proceedings of the Cambridge Philosophical Society 32}, 3 (1936), 477–481.

\bibitem{Perk2015TheEH}
{\sc Perk, J. H.~H.}
\newblock The early history of the integrable chiral potts model and the odd–even problem.
\newblock {\em Journal of Physics A: Mathematical and Theoretical 49\/} (2015).

\bibitem{pfeuty1970one}
{\sc Pfeuty, P.}
\newblock The one-dimensional ising model with a transverse field.
\newblock {\em ANNALS of Physics 57}, 1 (1970), 79--90.

\bibitem{Plischke1970}
{\sc Plischke, M., and Mattis, D.}
\newblock Two-dimensional ising model in a finite magnetic field.
\newblock {\em Phys. Rev. B 2\/} (Oct 1970), 2660--2663.

\bibitem{PronkoSyrygina2024}
{\sc Pronko, A.~G., and Syrygina, S.~K.}
\newblock Quantum {L}-operator of the critical {Ising} model.
\newblock In {\em Questions of quantum field theory and statistical physics. Part 30}, vol.~532 of {\em Zap. nauch. sem. POMI}. POMI, 2024, pp.~257--272.
\newblock Questions of quantum field theory and statistical physics, Part 30.

\bibitem{Sachdev_2011}
{\sc Sachdev, S.}
\newblock {\em Quantum Phase Transitions}, 2~ed.
\newblock Cambridge University Press, 2011.

\bibitem{Sato:1977uv}
{\sc Sato, M., Miwa, T., and Jimbo, M.}
\newblock {Studies on Holonomic Quantum Fields. 1.}
\newblock {\em Publ. Res. Inst. Math. Sci. Kyoto 14\/} (1978), 223--267.

\bibitem{Savit1981}
{\sc Savit, R.}
\newblock Duality in field theory and statistical systems.
\newblock {\em Rev. Mod. Phys. 52\/} (Apr 1980), 453--487.

\bibitem{Lieb1964}
{\sc Schultz, T.~D., Mattis, D.~C., and Lieb, E.~H.}
\newblock Two-dimensional ising model as a soluble problem of many fermions.
\newblock {\em Rev. Mod. Phys. 36\/} (Jul 1964), 856--871.

\bibitem{seiberg2024majorana}
{\sc Seiberg, N., and Shao, S.-H.}
\newblock Majorana chain and ising model-(non-invertible) translations, anomalies, and emanant symmetries.
\newblock {\em SciPost Physics 16}, 3 (2024), 064.

\bibitem{shao2023s}
{\sc Shao, S.-H.}
\newblock What's done cannot be undone: Tasi lectures on non-invertible symmetries.
\newblock {\em arXiv preprint arXiv:2308.00747\/} (2023).

\bibitem{Shastry1986ExactIO}
{\sc Shastry, B.~S.}
\newblock Exact integrability of the one-dimensional hubbard model.
\newblock {\em Physical review letters 56 23\/} (1986), 2453--2455.

\bibitem{Shastry1988DecoratedSR}
{\sc Shastry, B.~S.}
\newblock Decorated star triangle relations and exact integrability of the one dimensional hubbard model.
\newblock {\em Journal of Statistical Physics 50\/} (1988), 57--79.

\bibitem{Shiroishi1995YangBaxterEF}
{\sc Shiroishi, M., and Wadati, M.}
\newblock Yang-baxter equation for the r-matrix of the one-dimensional hubbard model.
\newblock {\em Journal of the Physical Society of Japan 64\/} (1995), 57--63.

\bibitem{singh2024unitarytetrahedronquantumgates}
{\sc Singh, V.~K., Sinha, A., Padmanabhan, P., and Korepin, V.~E.}
\newblock Unitary tetrahedron quantum gates.
\newblock {\em arXiv:2407.10731[quant-ph]\/} (2024).

\bibitem{Sklyanin1982QuantumVO}
{\sc Sklyanin, E.~K.}
\newblock Quantum version of the method of inverse scattering problem.
\newblock {\em Journal of Soviet Mathematics 19\/} (1982), 1546--1596.

\bibitem{Sklyanin1979QuantumIP}
{\sc Sklyanin, E.~K., Takhtadzhyan, L.~A., and Faddeev, L.~D.}
\newblock Quantum inverse problem method. i.
\newblock {\em Theoretical and Mathematical Physics 40\/} (1979), 688--706.

\bibitem{slavnov2019algebraicbetheansatz}
{\sc Slavnov, N.~A.}
\newblock Algebraic bethe ansatz.
\newblock {\em arXiv:1804.07350 [math-ph]\/} (2019).

\bibitem{Suzuki1971-1}
{\sc Suzuki, M.}
\newblock Relationship among exactly soluble models of critical phenomena. i*): 2d ising model, dimer problem and the generalized xy-model.
\newblock {\em Progress of Theoretical Physics 46}, 5 (11 1971), 1337--1359.

\bibitem{Suzuki1976-2}
{\sc Suzuki, M.}
\newblock Relationship between d-dimensional quantal spin systems and (d+1)-dimensional ising systems: Equivalence, critical exponents and systematic approximants of the partition function and spin correlations.
\newblock {\em Progress of Theoretical Physics 56}, 5 (11 1976), 1454--1469.

\bibitem{Takhtadzhan1979THEQM}
{\sc Takhtadzhan, L.~A., and Faddeev, L.~D.}
\newblock The quantum method of the inverse problem and the heisenberg xyz model.
\newblock {\em Russian Mathematical Surveys 34\/} (1979), 11--68.

\bibitem{Temperley:1971iq}
{\sc Temperley, H. N.~V., and Lieb, E.~H.}
\newblock {Relations between the 'percolation' and 'colouring' problem and other graph-theoretical problems associated with regular planar lattices: some exact results for the 'percolation' problem}.
\newblock {\em Proc. Roy. Soc. Lond. A 322\/} (1971), 251--280.

\bibitem{Vanicat2017IntegrableTL}
{\sc Vanicat, M., Zadnik, L., and Prosen, T.}
\newblock Integrable trotterization: Local conservation laws and boundary driving.
\newblock {\em Physical review letters 121 3\/} (2017), 030606.

\bibitem{Wannier1945}
{\sc Wannier, G.~H.}
\newblock The statistical problem in cooperative phenomena.
\newblock {\em Rev. Mod. Phys. 17\/} (Jan 1945), 50--60.

\bibitem{McCoy1976}
{\sc Wu, T.~T., McCoy, B.~M., Tracy, C.~A., and Barouch, E.}
\newblock Spin-spin correlation functions for the two-dimensional ising model: Exact theory in the scaling region.
\newblock {\em Phys. Rev. B 13\/} (Jan 1976), 316--374.

\bibitem{YangCN1967}
{\sc Yang, C.~N.}
\newblock Some exact results for the many-body problem in one dimension with repulsive delta-function interaction.
\newblock {\em Phys. Rev. Lett. 19\/} (Dec 1967), 1312--1315.

\bibitem{Yurov1991CORRELATIONFO}
{\sc Yurov, V.~P., and Zamolodchikov, A.~B.}
\newblock Correlation functions of integrable 2d models of the relativistic field theory: Ising model.
\newblock {\em International Journal of Modern Physics A 06\/} (1991), 3419--3440.

\bibitem{Zhang_2025}
{\sc Zhang, H.-C., and Sierra, G.}
\newblock Kramers-wannier self-duality and non-invertible translation symmetry in quantum chains: a wave-function perspective.
\newblock {\em Journal of High Energy Physics 2025}, 5 (May 2025).

\end{thebibliography}

\end{document}